\documentclass{aa}  

\usepackage{graphicx}
\usepackage{txfonts}
\usepackage{units}
\usepackage{soul}
\setstcolor{red}
\usepackage{keyval}
\usepackage{comment}
\usepackage{colortbl}
\usepackage{natbib}
\usepackage{hyperref} 
\usepackage{xcolor}
\usepackage{xspace}
\usepackage{subcaption}

\usepackage{tikz}
\definecolor{lime}{HTML}{A6CE39}
\DeclareRobustCommand{\orcidicon}{
    \begin{tikzpicture}
    \draw[lime, fill=lime] (0,0) 
    circle [radius=0.16] 
    node[white] {{\fontfamily{qag}\selectfont \tiny ID}};
    \draw[white, fill=white] (-0.0625,0.095) 
    circle [radius=0.007];
    \end{tikzpicture}
    \hspace{-2mm}
}

\newcommand{\orcidSAD}{\href{https://orcid.org/0000-0001-6010-6200}{\orcidicon}}
\newcommand{\orcidMK}{\href{https://orcid.org/0000-0002-5365-1267}{\orcidicon}}
\newcommand{\orcidJO}{\href{https://orcid.org/0000-0001-7776-498X}{\orcidicon}}
\newcommand{\orcidLL}{\href{https://orcid.org/0000-0002-5635-3345}{\orcidicon}}
\newcommand{\orcidGO}{\href{https://orcid.org/0000-0002-2863-676X}{\orcidicon}}
\newcommand{\orcidPG}{\href{https://orcid.org/0000-0003-2271-9297}{\orcidicon}}
\newcommand{\orcidJM}{\href{https://orcid.org/0000-0002-1963-6848}{\orcidicon}}
\newcommand{\orcidLF}{\href{https://orcid.org/0000-0003-2737-5681}{\orcidicon}}
\newcommand{\orcidAM}{\href{https://orcid.org/0000-0002-2564-3104}{\orcidicon}}
\newcommand{\orcidJF}{\href{https://orcid.org/0000-0001-8694-4966}{\orcidicon}}

\newcommand{\msun}{~M\mbox{$_\odot$}\xspace}
\newcommand{\gaia}{{\tt Gaia}\xspace}

\newcommand{\mpys}{{~mas\,yr$^{-2}$}\xspace}

\newcommand{\acca}{$\dot{\mu}_\alpha^*$\xspace}
\newcommand{\accd}{$\dot{\mu}_\delta$\xspace}

\defcitealias{kounkel2017}{KHL17}
\defcitealias{dzib2026}{Paper~I}

\begin{document}

\title{Dynamical masses of young stellar objects with the VLBA: DYNAMO-VLBA}
\subtitle{Radio binary stars in Orion}

\titlerunning{DYNAMO-VLBA: Binaries in Orion}

\author{
  Sergio A. Dzib\orcidSAD\inst{1} \and Jazm\'{\i}n Ord\'{o}\~nez-Toro\orcidJO\inst{2,3} \and Laurent Loinard\orcidLL\inst{2,4,5} \and 
  Marina Kounkel\orcidMK\inst{6} \and\\ Gisela N.\ Ortiz-Le\'on\orcidGO\inst{7} \and Phillip A. B. Galli\orcidPG\inst{8} \and 
  Luis F. Rodr\'{\i}guez\orcidLF\inst{2} \and Amy J. Mioduszewski\orcidAM\inst{9} \and\\ Josep M. Masqu\'e\orcidJM\inst{10,11}  \and Eoin O'Kelly\inst{12} \and Jan Forbrich\orcidJF\inst{12,13} \and Karla Moo-Herrera\inst{14}
}

\institute{
  \inst{1}{ Max-Planck-Institut f\"ur Radioastronomie, Auf dem H\"ugel 69, D-53121 Bonn, Germany;} \email{sdzib@mpifr-bonn.mpg.de}\\
  \inst{2}{ Instituto de Radioastronom\'{\i}a y Astrof\'{\i}sica, Universidad Nacional Aut\'onoma de M\'exico, Morelia 58089, M\'exico}\\
  \inst{3}{ Departamento de Astronom\'{\i}a, Universidad de Guanajuato, Apartado Postal 144, 36000 Guanajuato, M\'exico}\\
  \inst{4}{ Black Hole Initiative at Harvard University, 20 Garden Street, Cambridge, MA 02138, USA}\\
  \inst{5}{ David Rockefeller Center for Latin American Studies, Harvard University, 1730 Cambridge Street, Cambridge, MA 02138, USA}\\
  \inst{6}{ Department of Physics and Astronomy, University of North Florida, 1 UNF Dr., Jacksonville, FL, 32224}\\
  \inst{7}{ Instituto Nacional de Astrofísica, Óptica y Electrónica, Apartado Postal 51 y 216, 72000 Puebla, México}\\
 \inst{8}{ Instituto de Astronomia, Geofísica e Ciências Atmosféricas, Universidade de São Paulo, Rua do Matão, 1226, Cidade Universitária, 05508-090, São Paulo-SP, Brazil.}\\
 \inst{9}{ National Radio Astronomy Observatory, Domenici Science Operations Center, 1003 Lopezville Road, Socorro, NM 87801, USA}\\
 \inst{10}{ Departament de Física Quàntica i Astrofísica (FQA), Universitat de Barcelona (UB), c/ Martí i Franquès 1, 08028 Barcelona, Catalunya, Spain}\\
 \inst{11}{ Institut de Ciències del Cosmos (ICCUB), Universitat de Barcelona (UB), c/ Martí i Franquès 1, 08028 Barcelona, Catalunya, Spain}\\
 \inst{12}{ Centre for Astrophysics Research, University of Hertfordshire, College Lane, Hatfield, AL10 9AB, UK}\\
 \inst{13}{ Center for Astrophysics | Harvard \& Smithsonian, 60 Garden St, MS 72, Cambridge, MA 02138, USA}\\
 \inst{14}{ Facultad de Ingeniería, Universidad Aut\'onoma de Yucat\'an. Avenida Industrias No Contaminantes por Anillo Periférico Norte s/n, 97302, Mérida, Yucatán, México}\\
  }

\date{Received ; accepted }

\abstract
{
We present results from a multi-epoch Very Long Baseline Array (VLBA) survey conducted as part of the DYNAMO–VLBA project, aimed at measuring the dynamical masses of young stellar systems in the Orion complex. Our observations include 19 radio sources associated with 15 binary or multiple young systems. For four visual binaries in which both components were detected, the derived Keplerian orbits yield model-independent stellar masses; in particular, Brun~656 and HD~294300 show excellent agreement between VLBA-based and spectral-energy-distribution-based estimates, providing valuable benchmarks for pre-main-sequence evolutionary models. The component NU~Ori~C is confirmed as an intermediate-mass ($\sim$7\,M$_\odot$) star with nonthermal radio emission, offering rare evidence of magnetic activity near the boundary with the high-mass regime. Several additional sources exhibit astrometric accelerations or periodic residuals, revealing unseen companions and extending dynamical constraints to systems with only one radio-emitting component. These results highlight the capability of very long baseline interferometry astrometry to obtain precise and model-independent masses of young binaries, providing critical empirical anchors for stellar evolution models and new insights into the origin of magnetism in intermediate-mass stars.
}


\maketitle

\section{Introduction}
Binary and multiple stellar systems play a fundamental role in the star formation process and in the early dynamical evolution of stellar clusters.
Their frequency, separation distribution, and mass ratios provide essential tests for theories of core fragmentation and disk instability \citep[e.g.,][]{Duchene2013}.
In young star-forming regions, direct measurements of orbital motions are particularly valuable because they yield dynamical masses that can be compared with predictions from pre-main-sequence (PMS) evolutionary models.
Such measurements, however, are often challenging: optical spectroscopy provides only projected velocity amplitudes, adaptive-optics imaging offers limited resolution for close systems, and extinction frequently prevents optical access to the youngest members.

Radio interferometry provides a complementary avenue for studying young binaries.
Compact, nonthermal radio emission from magnetically active young stellar objects (YSOs) can be monitored at milliarcsecond (mas) scales with very long baseline interferometry (VLBI). Facilities such as the Very Long Baseline Array (VLBA) allow orbital motion to be traced directly over timescales of a few years. VLBI astrometry has already yielded accurate dynamical masses for nearby young multiple systems by directly tracing their stellar motions \citep[e.g.,][]{Loinard2007,Dzib2010,azulay2017,ortiz2017a,zhang2020}, demonstrating its unique capability to resolve orbits with stellar separations on the astronomical unit scale.

The Orion molecular complex, at a mean distance of $\sim$400\,pc \citep{kounkel2017,grosscheld2019}, provides a rich environment to extend such studies to a larger and more diverse population. It is composed of several nearby and well-studied regions of ongoing star formation, including the Orion Nebula Cluster (ONC) and NGC~2024 (see Fig.~\ref{fig:Obi1}). It hosts hundreds of nonthermal radio-emitting YSOs spanning a broad range of masses and evolutionary stages \citep{kounkel2017, forbrich2021}, making it ideal for investigating how multiplicity evolves with stellar mass and age. Within this framework, the {dynamical masses of multiple YSOs with the VLBA (DYNAMO-VLBA) project was designed to obtain multi-epoch VLBI observations aimed at resolving orbital motions and determining dynamical masses in nearby star-forming regions \citep{ordonez2024}, including the Orion complex, as well as $\rho$-Ophiuchus, Taurus, and Serpens. The program also records all compact radio sources within each observed field, allowing the measurement of parallaxes and proper motions for isolated objects. In \citet[][hereafter \citetalias{dzib2026}]{dzib2026}, we introduce the DYNAMO-VLBA observations of Orion, and present all detected sources, the astrometric fitting procedures, the derived astrometric parameters (including distances for individual sources and per region), and a detailed comparison of VLBA and \gaia\ astrometric results.

In this second paper of the DYNAMO-VLBA Orion series, we focus on the subsample of 15 binaries and multiple systems (see Table~\ref{tab:binaries} and Sect.~\ref{sec:binaries}). The observations, as well as astrometric and orbital fitting methodology adopted in this work are described in Sect.~\ref{sec:data}, following the same framework introduced in Paper~I. By combining new VLBA observations with earlier VLBA epochs from the Gould's Belt Distances Survey (GOBELINS; \citealt{ortiz2017a} and \citealt{kounkel2017}, the latter hereafter \citetalias{kounkel2017}), we determined full or partial orbits for these systems and derived total and, when possible, individual stellar masses. Our results (Sect.~\ref{sec:results}) provide a homogeneous set of VLBI-based dynamical masses for young binary and multiple systems in the Orion molecular cloud complex, derived using a uniform astrometric methodology.
In Sect.~\ref{sec:discussion}, we discuss the implications of these results, including the occurrence of compact, nonthermal radio emission in intermediate-mass systems, and in Sect.~\ref{sec:conclusions} we summarize our main findings and conclusions.

\begin{table*}
\scriptsize
\setlength{\tabcolsep}{4pt}
\renewcommand{\arraystretch}{1.1}
\begin{minipage}{\linewidth}
\caption{The DYNAMO-VLBA project targeted binaries in Orion.}\label{tab:binaries}
\begin{tabular}{ccccccccccccccccccc}
\hline\hline
     &  VLBA   &   VLBA  &              Binary      & Spectral  & YSO   &    Mass 1 & Mass 2 & Mass 3   & Ref. & Orion\\
Name &  Name 1 & Name 2  &              type        &  type     & class &(M$_\odot$)&(M$_\odot$)& (M$_\odot$)&     &Region\\
\hline
CXOU J054146.1-015622    & 61  & 62 & VLBI visual           & ...     & YSO   &  $1.85\pm0.58$&$0.95\pm0.22$&  ...   & 1, 2&NGC 2024\\
Brun 656                 & 107 &  4  & VLBI visual           & G2III   & III   &  {2.72$^{+0.21}_{-0.15}$}&...  & ... & 2, 3 & ONC\\
HD 294300$^{\rm a}$      &145 & 145B &   VLBI visual           & G1     & III   &  2.47$^{+0.15}_{-0.17}$&...&  ...   & 2, 10, 11 & $\sigma$ Orionis\\ 
2MASS J05414134-0153326$^{\rm b}$  &153 & 153B & VLBI visual & ...& I/II&...&...&...&12   &NGC\,2024\\
V* NU Ori                & 27  & ...     & Hierarchical/Spectroscopic &B0.5V&Massive&$14.9\pm0.5$&$3.9\pm0.7$& $7.8\pm0.7$&4, 5 & ONC \\
$\theta^1$\,Ori A  & 11 & ... &  Hierarchical/Spectroscopic& B0.5+A0V+A? & Massive & 15.3 & 4.0 & 2.5 &17,18,19 & ONC \\
$\theta^1$\,Ori E  & 9 & ... & Spectroscopic & G2IV & ... & 2.807 & 2.797 & ... &20, 21 & ONC \\
Parenago 1540$^{\rm c}$  & 19 & ... & Spectroscopic& K3V+K5V& WTTS & $1.7$& $1.25$&... &13, 14& ONC\\
HD\,37017$^{\rm c}$  & 34 & ... & Spectroscopic& B2V&Massive & 3.5 to 8.5 & 1.8 to 4.5 & ... &15, 16 & ONC\\
COUP 450 [GMR A]         & 5 &  ...      & Astrometric             & K5V     &III   & ...&...&...&  6,7,8 & ONC\\
ALMA J053514.5010-052238.674& 7& ...     & Astrometric             & ...     &...   & ...&...&...&  8 & ONC\\
V* MT Ori                & 8   & ...     & Astrometric             & K2      &...   & ...&...&...&  2, 8, 22 & ONC\\ 
V* V1399 Ori$^{\rm d}$             &GMR V  & ... & Astrometric  & G8 & III& ... &...& ... &8, 9&ONC\\ 
CXOU J054145.8-015411    &124 & ... &  Astrometric  &... & YSO&...&...&...&1, 12& NGC\,2024 \\
V* V1230 Ori             &149   &  ... &  Astrometric  & B1 & Massive& ...&...&...&1& ONC \\
\hline\hline
\end{tabular}
\end{minipage}
\tablefoot{\scriptsize $^{\rm a}$ \citetalias{kounkel2017} only reported VLBA 145 and labeled as a probable binary. We have detected the companion in new VLBA observations. \\
$^{\rm b}$ No previous suggestion of binarity prior to this work. However, our new VLBA observations uncovered a companion to this source. \\
$^{\rm c}$ Not included in DYNAMO-VLBA observations; however, we did a new astrometric analysis and include it here for completeness.\\
$^{\rm d}$ Not included in early VLBA observations of GOBELINS. \\
\\References are: 1 = \citet{broos2013}, 2 = \citetalias{kounkel2017}, 3 = \citet{valegrad2021}, 4 = \citet{simon2011},
5 = \citet{schultz2019},  6 = \citet{Bower2003}, 7 = \citep{grosscheld2019}, 8 = \citep{dzib2021}, 9 = \citet{hillenbrand1997}, 10 = \citet{pinzon2021}, 11 = \citet{hernandez2014}, 12 = this work, 13 = \citet{marschall1988}, 14 = \citet{palla2001}, 15\,=\,\citet{bolton1998}, 16 = \citet{hohle2010}, 17 = \citet{lloyd1999}, 18 = \citet{petr1998}, 19 = \citet{schertel2003}, 20 = \citet{costero2006}, 21\,=\,\citet{morales2012}, and 22\,=\,\citet{jeffries2007}.
}
\end{table*}

\section{Binaries in Orion studied within DYNAMO-VLBA}\label{sec:binaries}

The DYNAMO-VLBA survey has targeted a sample of young stars in the Orion complex to investigate multiplicity at milliarcsecond scales. This section summarizes the main binary and multiple systems identified or characterized in the course of the program. The systems span a wide range of stellar masses and evolutionary stages, from low-mass embedded YSOs to massive OB-type stars, and include both previously known binaries and new candidate multiples. In Table~\ref{tab:binaries} we compile previous identifications, spectral classifications, and mass estimates. We briefly discuss evidence for multiplicity based on earlier observations.

\paragraph{VLBI visual binaries.} 
These are systems in which both stellar components are directly detected in VLBA images. They are of particular importance within the DYNAMO–VLBA survey, since for those with sufficient orbital coverage the VLBA data allow a direct determination of the individual stellar masses.  
Within this category we identify four systems. CXOU J054146.1--015622 is a deeply embedded young star that has only been detected in X-rays and radio wavelengths \citep{skinner2003,rodriguez2003,broos2013}. \citetalias{kounkel2017} reported two compact radio components (VLBA 61 and 62) separated by $\sim$15\,mas, and provided preliminary mass estimates for both components. The young star Brun\,656 \citep{walker1969} is located in the Orion Nebula Cluster (ONC) and in which \citetalias{kounkel2017} detected two radio sources (VLBA 4 and 107) and provided lower-limit mass estimates of $\sim$1.7\msun\ for both components, suggesting an intermediate-mass system. Later, \citet{valegrad2021} performed a detailed spectral energy distribution (SED) analysis and estimated a stellar mass of $2.72^{+0.21}_{-0.15}$\msun, proposing that this system may be a precursor to a Herbig Ae/Be star.
Finally, while \citetalias{kounkel2017} reported single radio sources associated with HD~294300 and 2MASS~J05414134$-$0153326, our new VLBA observations reveal the companions to both, confirming their binary nature.

\paragraph{Spectroscopic binaries.} 
A second group within our sample consists of spectroscopic binaries previously identified at optical wavelengths and now detected as compact radio sources with the VLBA. Because the spectroscopic mass ratios and orbital parameters are known for these systems, individual stellar masses can, in principle, be derived by combining the spectroscopic solutions with the VLBA astrometry, a possibility already explored by \citetalias{kounkel2017}.  
HD\,37017 (V* V1046 Ori, VLBA 34) and Parenago\,1540 (VLBA 19) are well-established spectroscopic binaries with orbital periods of $18.6561\pm0.0002$ and $33.73\pm0.03$ days, respectively \citep{bolton1998,marschall1988}. Both were detected as radio sources by \citetalias{kounkel2017}, who provided preliminary component mass estimates.  
$\theta^1$\,Ori\,E (VLBA\,9) is another short-period spectroscopic binary ($P \approx 9.9$ days), also eclipsing, with accurately measured masses of $2.807$ and $2.797$\,M$_\odot$ \citep{costero2006,morales2012}.

\paragraph{Hierarchical triple and higher-order systems.}
These are systems where more than two stars are gravitationally bound. 
V*\,NU\,Ori (HD\,37061) is a B0.5V star and the primary ionizing source of the M43 H\,\textsc{ii} region \citep{simon2011}. It is a known hierarchical quadruple system. At optical wavelengths, it appears as a visual binary (components A and B) with a separation of $0\rlap{.}''47$ \citep{preibisch1999}. Component A is itself a spectroscopic triple system: the primary Aa and its close companion Ab orbit each other every 14 days, while a third component (C) orbits the Aa+Ab binary with a period of 467 days. From spectroscopic and evolutionary model analysis, \citet{schultz2019} derived component masses of $M_{\rm Aa}=14.9\pm0.5$\msun, $M_{\rm Ab}=3.9\pm0.7$\msun, and $M_{\rm C}=7.8\pm0.7$\msun. No mass has yet been derived for V*\,NU\,Ori\,B. All known components are of intermediate or high mass. V*\,NU\,Ori was also suggested to be a magnetic active star; the magnetic fields were initially attributed to the primary \citep{petit2008}, but \citet{schultz2019} showed it is actually associated with component C. \citetalias{kounkel2017} detected two radio sources toward this system, VLBA 27 and 28, but with only 3 and 2 detections respectively, no astrometric solution was derived. Our reanalysis indicates that one detection previously labeled as VLBA\,28 likely corresponds to VLBA\,27, and the other is likely a spurious detection, leaving a single confirmed radio component associated with V*\,NU~Ori. Finally, $\theta^1$\,Ori\,A is a hierarchical triple system, composed of a spectroscopic binary ($\theta^1$\,Ori\,A$_1$ and $\theta^1$\,Ori\,A$_3$) and a wide companion ($\theta^1$\,Ori\,A$_2$). While \citetalias{kounkel2017} associated their radio source VLBA\,11 with the spectroscopic pair A$_1$+A$_3$, \citet{dzib2021} showed it instead corresponds to the wider companion A$_2$, a $\sim4$\,M$_\odot$ star at $\sim0\rlap{.}''18$ separation \citep{petr1998,schertel2003,gravity2018}.

\paragraph{Suspected astrometric binaries.}
Several embedded Orion stars show nonlinear or anomalous motions in VLBA data, indicating the presence of unresolved companions \citep[\citetalias{kounkel2017};][]{dzib2021}. These include COUP\,450 (GMR\,A, VLBA\,5), ALMA\,J053514.5010--052238.674 (COUP\,639, GMR\,H, VLBA\,7), V*\,MT\,Ori (GMR\,G,\,VLBA\,8), and V*\,V1399\,Ori (GMR\,V).  In addition, 2MASS~J05414134$-$0153326 (VLBA\,153), V*\,V1230\,Ori (VLBA\,149), and CXOU~J054145.8$-$015411 (VLBA\,124), which had only two detections in \citetalias{kounkel2017} and were not previously analyzed astrometrically, show positional deviations in our new data consistent with binarity.  
Although only one component is detected at radio wavelengths, and therefore individual stellar masses cannot be determined, a description of its motion can put constraints on the system properties, as discussed in the next section.

\section{Data and analysis methods}\label{sec:data}
\subsection{VLBA observations}
The observations analyzed in this paper are part of the DYNAMO-VLBA project and were obtained with the VLBA under project code BD215.  
The full observational setup, calibration procedures, and astrometric analysis have been described in detail in \citetalias{dzib2026}.
Here we summarize only the main aspects relevant to the binary and multiple systems.

Each of the seven DYNAMO-VLBA Orion pointings (blocks F-L, see \citetalias{dzib2026}) was observed at a frequency of 5.0\,GHz with a total bandwidth of 256\,MHz.  
Between February~2018 and December~2019, six epochs were obtained per block, separated by 2 to 6~months depending on the expected orbital periods of the known binaries.  
Additional earlier epochs from the GOBELINS project \citep{kounkel2017} were included when available, extending the temporal baseline back to 2016.

Data calibration and imaging were carried out in \textsc{AIPS} following the procedures described in \citetalias{dzib2026}. These include the standard calibration scheme, multiphase calibrator referencing, and geodetic block corrections.  
The mean synthesized beam was $2.5\times1.0$\,mas, and the typical image rms noise was about 30\,$\mu$Jy\,beam$^{-1}$.  
Astrometric positions for all detected components were obtained through two-dimensional Gaussian fitting with \textsc{JMFIT}.

From the 216 compact radio sources detected in the full DYNAMO-VLBA+GOBELINS Orion dataset, 19 were identified to be related to 15 binaries or higher–order multiples showing measurable positional changes inconsistent with linear motion.  
These systems constitute the sample analyzed in the present paper.  
Their selection was based on (i) multiple detections of at least two compact components, or (ii) significant curvature or acceleration in the single–source astrometric fits from \citetalias{dzib2026}.  
For each system, we combined the positions of the individual components across all available epochs to model their relative and absolute motions.

\subsection{Motion fitting and mass estimation}

The astrometric positions ($\alpha,\,\delta$) of all detected components were modeled following the same formalism described in \citetalias{dzib2026}. In this approach, each trajectory is represented as the combination of trigonometric parallax ($\varpi$) and linear proper motions ($\mu_\alpha\cdot\cos{(\delta)},\,\mu_\delta$).  
For binary and multiple systems, the model was extended as required by the data to include either additional acceleration terms (for long-period orbits, only partially covered by the VLBA time baseline) or full Keplerian orbital terms describing the absolute motion of the detected component(s). The equations used to describe these motions are

\begin{align}
\alpha(t) &= \alpha_0 + \mu_\alpha \cdot \cos(\delta) \cdot t + \varpi \cdot f_\alpha(t) + \Delta\alpha_{orb}(t), \\
\delta(t) &= \delta_0 + \mu_\delta \cdot t + \varpi \cdot f_\delta(t) + \Delta\delta_{orb}(t),
\end{align}

\noindent where $(f_\alpha, f_\delta)$ represent the parallactic factors projected along right ascension and declination. $\Delta\alpha_{\mathrm{orb}}(t)$ and $\Delta\delta_{\mathrm{orb}}(t)$ represent the orbital contribution induced by the presence of a companion.  
These terms may correspond to the contribution of a constant-acceleration approximation ($\dot{\mu}_\alpha\cdot\cos{(\delta)}$, $\dot{\mu}_\delta$) for systems with long orbital periods, or to the full Keplerian description. The latter case is defined by the orbital elements: Period ($P$), semi-major axis ($a$), eccentricity ($e$), orbit inclination ($i$), the argument of ascending node ($\Omega$), the argument of pericenter ($\omega$), and the time of passage of pericenter ($T_0$).

For systems in which both components are detected at radio wavelengths, the relative orbit and the absolute motions of each star can be jointly fitted.  
The combination of the orbital semi-major axes ($a_1$, $a_2$) and the period ($P$) then yields the total and individual stellar masses through Kepler’s third law,  
$M_1 + M_2 = (a_1 + a_2)^3 / P^2$, and $M_1/M_2 = a_2/a_1$.  
This approach provides model-independent dynamical masses when the orbital coverage is sufficient.

For systems with only one radio-detected component, useful dynamical information can still be derived depending on the type of binary.  
In spectroscopic binaries, where the mass ratio ($q=M_1/M_2$) and some orbital parameters are known from optical spectroscopy, these quantities were used as priors in the VLBA modeling, allowing individual masses to be inferred.  
In sources showing periodic astrometric motion but lacking spectroscopic information, the fitted orbital curvature yields an {astrometric mass function}, defined as

\begin{equation}\label{eq:3}
f(M) = \frac{M_2^3 \sin^3 i}{(M_1 + M_2)^2} = \frac{a_1^3}{P^2},
\end{equation}

\noindent where $a_1$ is the semi-major axis of the detected component (in~AU) and $P$ is the orbital period (in~yr).  
The mass function provides a lower limit on the companion mass and, when combined with an estimate of the primary mass or inclination, allows further constraints on the total system mass.  
 
Finally, in long-period binaries characterized only by measurable accelerations, the observed curvature in the trajectory can be used to constrain the total system mass.  
All results from the fits are presented in Sect.~\ref{sec:results} and discussed in Sect.~\ref{sec:discussion}.

\section{Results} \label{sec:results}

\subsection{Overview of the astrometric solutions}

The parallaxes and proper motions of all compact radio sources in the DYNAMO–VLBA survey, including the targets analyzed here, are presented in \citetalias{dzib2026}.  In this paper, we focus on the subset of sources that are known members of multiple systems or that exhibit clear signatures of binarity, such as orbital motion, measurable accelerations, or periodic residuals in position.  For these objects, we derive and report the corresponding orbital (Table~\ref{tab:dm}) or acceleration parameters. From the orbital parameters we estimate either the individual stellar masses or the astrometric mass function (Table~\ref{tab:mass}). 

From the 15 analyzed systems, we derived orbital parameters for 11 and acceleration parameters for two. Figure~\ref{fig:Obi1}
shows the orbital or acceleration solutions obtained for these 13 systems. The motions of the two remaining sources are consistent with linear trajectories affected only by the trigonometric parallax. The best-fit models to the positions of all 15 systems are shown in Fig.~\ref{fig:Mot}, and a detailed description of each case is provided below.

\begin{figure*}
    \centering
    \includegraphics[width=0.9\linewidth]{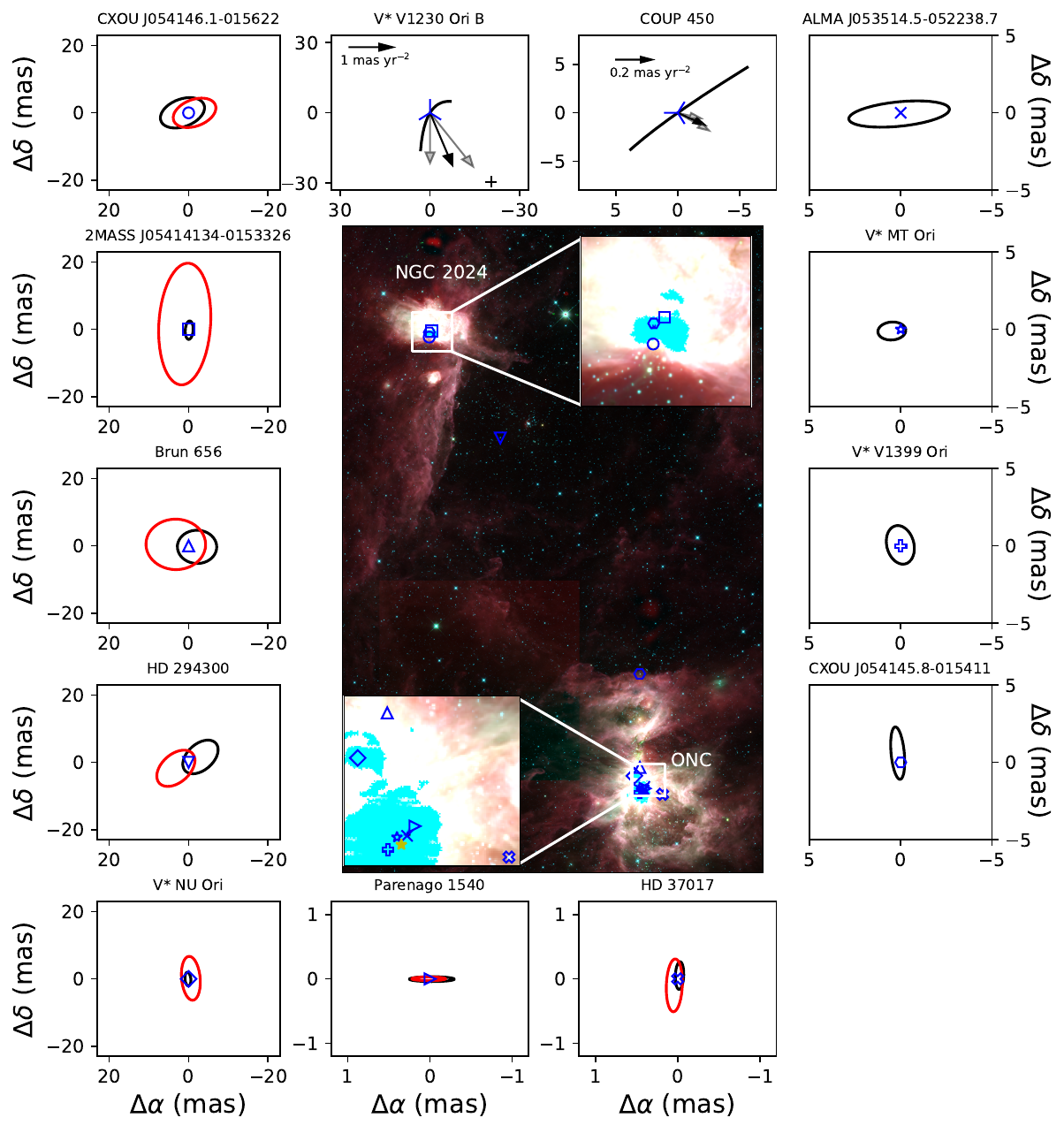}
    \caption{Orion binaries studied in DYNAMO-VLBA for which orbital motion or significant astrometric acceleration could be derived. {Central panel:} Wide-field infrared view of the Orion star-forming region using WISE imaging. RGB image is composed from W3 (12\,$\mu$m, red), W2 (4.6\,$\mu$m, green), and W1 (3.4\,$\mu$m, blue) bands. Cyan regions indicate areas affected by saturation in the \emph{WISE} images. Blue symbols mark the positions of 13 binaries observed in our sample, where orbital or acceleration parameters could be derived. Marker shapes are unique to each system for clarity. Zoomed views of the ONC and NGC\,2024 regions are also shown. The yellow star marks the position of $\theta_1$\,Ori\,C for reference.  The {surrounding subplots} display the projected stellar orbital motions, plotted as apparent motion on the sky.  In panels showing two ellipses, the black and red curves correspond to the best-fit astrometric models for the primary and secondary components, respectively. In panels with a single ellipse, the curve represents the orbital motion of the detected radio source. In the cases of V*\,V1230\,Ori\,B and COUP\,450, the arrows represent the measured acceleration vectors and errors. For V*\,V1230\,Ori\,B, the black cross indicates the relative position of the associated \gaia source, likely corresponding to the primary component of the system.
    } 
    \label{fig:Obi1}
\end{figure*}

\begin{table*}
\scriptsize
\setlength{\tabcolsep}{1.3pt}
\renewcommand{\arraystretch}{0.95}
\begin{minipage}{\linewidth}
\caption{Orbital parameters of radio sources related to visual, spectroscopic, and astrometric binaries (Sects. \ref{sec:visual} -- \ref{sec:astrome}). }\label{tab:dm}
\vspace{-0.2cm}
\begin{tabular}{ccccccccccccccccccc}
\hline\hline
 &  VLBA   &   VLBA      & $P$ & $T_P$& $e$ & $\omega$ & $i$ & $\Omega$ & $q= M_2/M_1$ & $a_1$ & $a_2$ \\
Name&Name 1 & Name 2 &(yr) & (JD) &  & ($^\circ$) & ($^\circ$) & ($^\circ$) &  & (AU) & (AU) \\
\hline
CXOU J054146.1--015622& 61 &  62 & $6.69 \pm 0.25$ & $2461021.5 \pm 110.9$ & $0.295 \pm 0.034$ & $316.7 \pm 8.4$ & $132.1 \pm 4.8$ & $147.7 \pm 3.9$ & $1.03 \pm 0.06$ & $2.36 \pm 0.11$ & $2.28 \pm 0.02$ \\
Brun 656 &107 & 4 & $6.29 \pm 0.12$ & $2460102.5 \pm 44.7$ & $0.434 \pm 0.027$ & $272.0 \pm 17.4$ & $25.5 \pm 6.6$ & $95.6 \pm 19.0$ & $0.66 \pm 0.03$ & $2.21 \pm 0.03$ & $3.36 \pm 0.09$ \\
HD 294300&145 & 145B & $6.57 \pm 0.55$ & $2454853.9 \pm 49.3$ & $0.672 \pm 0.050$ & $235.4 \pm 11.1$ & $51.7 \pm 5.6$ & $109.4 \pm 6.2$ & $0.93 \pm 0.16$ & $2.65 \pm 0.40$ & $2.86 \pm 0.06$  \\ 
2MASS J05414134-0153326&153 & 153B & $18.03 \pm 0.00$ & $2460619.8 \pm 127.3$ & $0.183 \pm 0.107$ & $63.0 \pm 4.7$ & $110.9 \pm 2.9$ & $93.2 \pm 1.3$ & $0.15 \pm 0.03$ & $1.12 \pm 0.19$ & $7.36 \pm 0.07$  \\
V* NU Ori & 27 &...  & $1.30 \pm 0.01$ & $2456026.5 \pm 13.7$ & $0.246 \pm 0.050$ & $94.9 \pm 8.4$ & $110.4 \pm 4.1$ & $85.6 \pm 4.2$ & $3.20$ & $2.76 \pm 0.03$ & $0.88 \pm 0.01$  \\
Parenago\,1540&19 & ... & $0.09$ & $2454501.5$ & $0.120$ & $131.3$ & $94 \pm 34$ & $177 \pm 27$ & $1.32 $ & $0.107\pm0.004$ & $0.081\pm0.004$   \\
HD 37017&34 & ... & $0.05$ & $2458853.5$ & $0.468$ & $118.3$ & $107 \pm 50$ & $90 \pm20$ & $0.52 $ & $0.072\pm0.001 $ & $0.137\pm0.001 $  \\
%
%
ALMA J053514.5010-052238.674& 7 & ... & $4.98\pm 0.17$ & $2456305.6 \pm 276.0$ & $0.092 \pm 0.051$ & $89.0 \pm 45.9$ & $106.3 \pm 1.0$ & $173.1 \pm 1.1$ & ... & $1.04 \pm 0.03$ &...  \\
V* MT Ori&8 & ... & $4.81 \pm 0.59$ & $2456198.01 \pm 237.59$ & $0.669 \pm 0.156$ & $164.3 \pm 118.3$ & $26.8 \pm 31.8$ & $74.7 \pm 115.0$ & ... & $0.30 \pm 0.06$ & ...  \\ 
V* V1399 Ori&GMR V & ... & $1.99 \pm 0.04$ & $2456957.46 \pm 330.31$ & $0.093 \pm 0.157$ & $171.6 \pm 157.2$ & $54.9 \pm 6.3$ & $79.4 \pm 10.4$ & ... & $0.52 \pm 0.03$ &... \\ 
CXOU J054145.8--015411&124 & ... & $5.06 \pm 0.27$ & $2455847.9 \pm 226.3$ & $0.474 \pm 0.116$ & $222.4 \pm 27.0$ & $102.7 \pm 7.4$ & $87.7 \pm 11.2$ & ... & $0.78 \pm 0.10$ & ...  \\
\hline\hline
\end{tabular}
\tablefoot{Columns from left to right are: The system common name, VLBA names (following the \citetalias{kounkel2017} notation when available, otherwise we used the common name given at radio wavelengths) for the first and second associated radio sources, orbital period ($P$), the time of passage of pericenter ($T_P$) eccentricity ($e$),  the argument of pericenter ($\omega$), orbit inclination ($i$), the argument of ascending node ($\Omega$), mass ratio ($q$), and semi-major axes of first ($a_1$) and second component ($a_2$).\\
Orbital and mass ratio parameters without given errors indicate that these were taken from the literature. }
\end{minipage}
\end{table*}

\begin{table}
\scriptsize
\setlength{\tabcolsep}{1.3pt}
\renewcommand{\arraystretch}{0.95}
\begin{minipage}{\linewidth}
\caption{Dynamical masses and astrometric mass functions, $f(M)$, of radio sources related to visual, spectroscopic, and astrometric binaries (Sects. \ref{sec:visual} -- \ref{sec:astrome}). }\label{tab:mass}
\vspace{-0.2cm}
\begin{tabular}{ccccccccccccccccccc}
\hline\hline
 &  $M_{\rm Total}$ & $M_1$ & $M_2$  & $f(M)$\\
Name& (M$_\sun$) & (M$_\sun$) & (M$_\sun$)&  (\msun)\\
\hline
CXOU J054146.1--015622& $2.24 \pm 0.04$ & $1.10 \pm 0.01$ & $1.14 \pm 0.05$ &...\\
Brun 656 &  $4.38 \pm 0.29$ & $2.64 \pm 0.13$ & $1.74 \pm 0.16$ &...\\
HD 294300& $3.89 \pm 0.08$ & $2.01 \pm 0.12$ & $1.87 \pm 0.21$ &...\\ 
2MASS J05414134-0153326& $1.92 \pm 0.16$ & $1.66 \pm 0.10$ & $0.26 \pm 0.06$&... \\
V* NU Ori & $28.64 \pm 1.30$ & $6.96 \pm 0.32$ & $21.68 \pm 0.99$ &...\\
Parenago\,1540& $0.78 \pm 0.10$ & $0.34 \pm 0.04$ & $0.44\pm 0.05$ &...\\
HD 37017& $3.48 \pm 0.10$ & $2.28 \pm 0.07$ & $1.20 \pm 0.03$ &...\\
ALMA J053514.5010-052238.674& ...& ... & ... & $0.045\pm0.005$\\
V* MT Ori&...& ... & ... & $0.0011\pm0.0008$ \\ 
V* V1399 Ori&...& ... & ... & $0.036\pm0.007$ \\ 
CXOU J054145.8--015411& ...& ... & ... & $0.018\pm0.007$\\
\hline\hline
\end{tabular}
\end{minipage}
\end{table}

\subsection{Radio sources related to multiple YSOs}

Binary and multiple systems constitute a significant fraction of the radio-detected YSOs in Orion. In total, we identify 19 sources associated with multiple systems, which we classify into four categories based on the nature of the detections and orbital constraints: (i) {radio visual binaries}, where two radio components are resolved and full orbital solutions can be derived; (ii) {spectroscopic binaries}, where external spectroscopic constraints on mass ratios and/or orbital elements are combined with our astrometry; (iii) {astrometric binaries}, identified from sinusoidal residuals in the positional fits of single detected components; and (iv) {wide binaries}, for which orbital periods are much longer than the VLBA time baseline and binarity manifests primarily as constant accelerations rather than closed orbits.

For categories (i)–(iii), we apply the orbital fitting procedures described in \citetalias{dzib2026}, and the reported reference positions in that paper correspond to the system barycenter rather than the locations of individual radio components at the reference epoch. The resulting orbital parameters and dynamical masses are listed in Tables~\ref{tab:dm} and ~\ref{tab:mass}. For category (iv), no orbital solution is attempted; instead, as outlined in \citetalias{dzib2026}, we model the motion as linear plus constant acceleration to capture the observed curvature over the available time span. In the following subsections, we present the results for each category and discuss their astrometric signatures and implications for dynamical mass estimates.

\begin{figure*}
    \centering
    \includegraphics[width=0.97\linewidth]{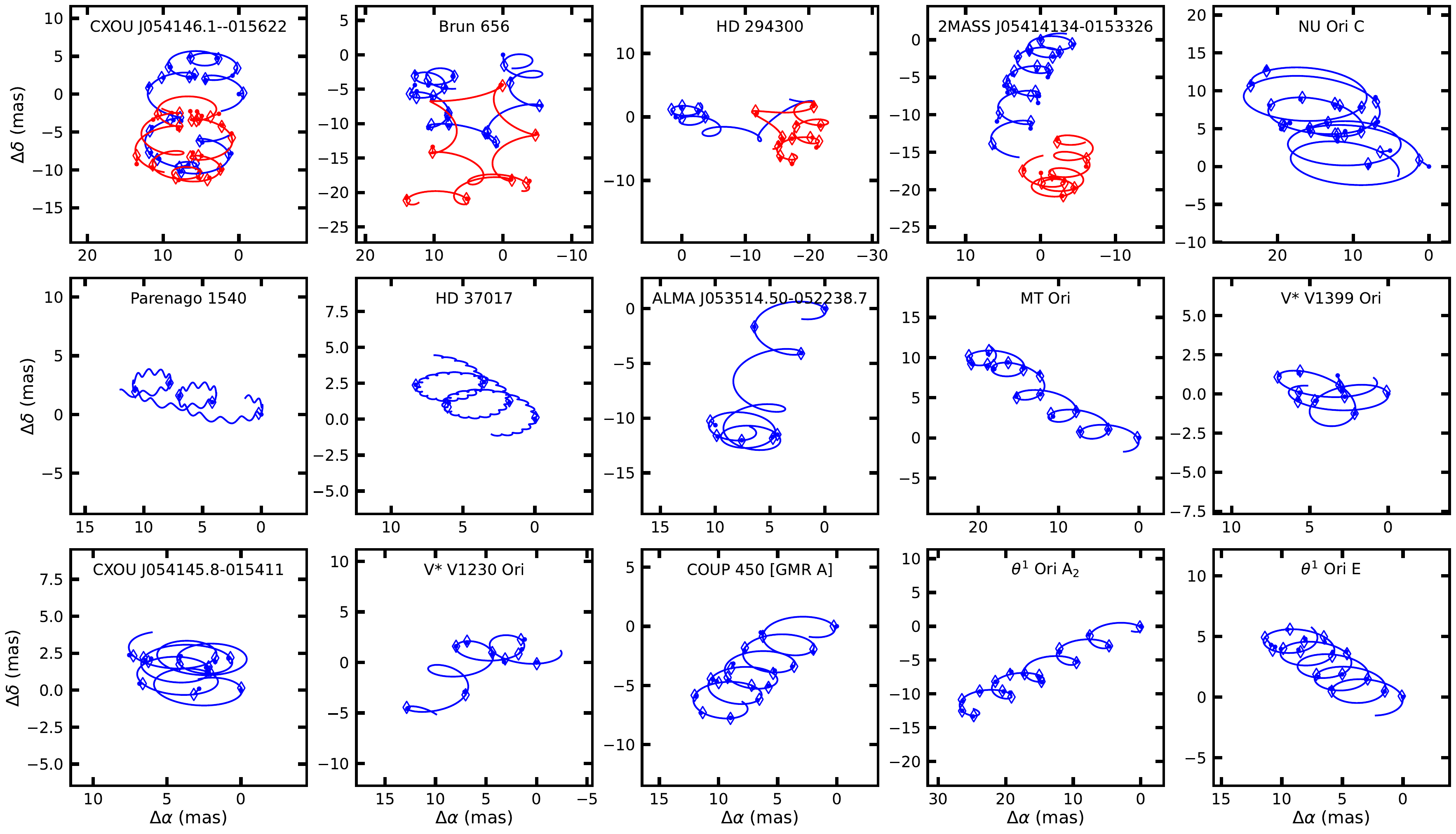}
    \caption{Best-fit astrometric motions of the radio sources associated with stellar multiple systems. Filled circles mark the measured positions relative to the first detection, solid lines show the best-fit trajectories, and open diamonds indicate the expected positions from the fitted model at the corresponding epochs. Different colors represent the individual stellar components detected within each system. 
    } 
    \label{fig:Mot}
\end{figure*}

\subsubsection{Radio visual binary stars}\label{sec:visual}

Visual binaries resolved in our VLBA observations provide the most direct means of determining dynamical masses, as the orbital motions of both components can be traced without additional assumptions. When two radio sources are detected in the same system with separations of a few tens of mas, their relative motion around the common center of mass yields strong constraints on orbital parameters and stellar masses \citep[e.g., ][]{ordonez2024}.

We identified four such systems: CXOU\,J054146.1--015622, Brun\,656, HD\,294300, and 2MASS\,J05414134--0153326 (see also their corresponding panels in Fig.~\ref{fig:Mot}). In the latter two cases, we report for the first time the detection of the companion radio sources to those previously identified by \citetalias{kounkel2017}, which we designate VLBA\,145B and VLBA\,153B, respectively. For these visual binaries, both stellar orbits are well sampled, and the resulting dynamical masses are highly robust, as they follow directly from the measured motions without requiring external constraints.

\subsubsection{Spectroscopic binaries}\label{sec:spectro}

Spectroscopic binaries provide a unique advantage for astrometric studies, as independent spectroscopic measurements constrain key orbital parameters such as mass ratios and, in some cases, orbital elements. When combined with VLBA astrometry, these constraints enable more robust orbital fits and improved stellar mass determinations, even if only one radio component is detected.

Four of our targets (V*\,NU\,Ori, $\theta_1$\,Ori\,E, Parenago\,1540, and HD\,37017) are known spectroscopic binaries. In all cases, only one radio source was detected in our images. We therefore adopted the mass ratios derived from spectroscopy and, where available, fixed additional orbital parameters to their known values, following the procedure outlined by \citetalias{kounkel2017}. This approach allows us to estimate the masses of the detected radio star and its unseen companion. 

In the case of V*\,NU\,Ori, the best astrometric fit was obtained by associating the detected radio source with component C, which orbits the close pair Aa and Ab. The identification is also consistent with previous suggestions that component C is the magnetically active star. The mass ratio of V*\,NU\,Ori was fixed to ${\rm M}_{\rm Aa+Ab}/M_{\rm C}=(14.9+3.9)/7.8=3.2$. For Parenago\,1540 or HD\,37017, no additional VLBA observations were obtained, but we carried out new fits to determine their barycentric positions at the reference epoch 2016.0, which had not been reported previously. For Parenago\,1540  our orbital and mass estimates are in good agreement with those of \citetalias{kounkel2017}, whereas for HD\,37017 we obtained lower masses. In both cases, our error bars are also smaller than those of \citetalias{kounkel2017}; however, this likely reflects the fact that we fixed orbital parameters from the literature to their most probable values, without propagating their uncertainties. Finally, the tightness of $\theta_1$\,Ori\,E system and its short orbital period ($\sim$9.9 days), prevents us from deriving reliable orbital and mass parameters, and we therefore treat it as a single star in our astrometric analysis and following analysis.

\subsubsection{Astrometric binaries}\label{sec:astrome}

Astrometric binaries are systems in which only one component is detected, but the orbital motion can be inferred from deviations in the measured position relative to a simple linear motion plus parallax model. In such cases, periodic residuals provide strong evidence for binarity, even when the companion is not directly observed.

In our sample, six radio sources (ALMA J053514.5010--052238.674, V* MT Ori, V* V1399 Ori, and CXOU J054145.8--015411) display significantly larger astrometric residuals than typical single stars, consistent with earlier classifications of these objects as binary candidates. The residuals exhibit sinusoidal patterns, suggesting orbital periods comparable to the observational time baseline. For these systems, we therefore modeled the motion by including full orbital parameters, despite the absence of a directly detected companion. The orbital period itself remains uncertain, but we constrained our fits to periods shorter than 10 years, guided by the variability observed in the astrometric residuals.

It is important to emphasize that the orbital solutions derived for these astrometric binaries should be treated with caution. Since only one component is detected, the fits rely on indirect signatures of binarity, which can lead to underestimated uncertainties in the derived parameters. As a result, while the fitted orbits are consistent with the observed residuals, the true parameter uncertainties are likely larger than those formally reported by the model.

\subsubsection{Wide binaries}

Wide binary systems, with orbital periods much longer than the VLBA time baseline, require a different treatment than close binaries. In these cases, only partial orbital coverage is available, and binarity signatures appear primarily as constant accelerations in the astrometric fits rather than complete orbital solutions.

In the case of source $\theta_1$\,Ori\,A$_2$ the companion at $\sim0\rlap{.}''18$ ($\sim$70 AU) from $\theta_1$\,Ori\,A$_1$ we proceed differently. As the total mass of the system is $\sim20$\,\msun, the period is estimated to be $\sim130$ years, significantly longer than the $\sim5$ years covered by the VLBA observations used in this work. For this system, we performed an astrometric fit that includes linear motion with a constant acceleration (see \citetalias{dzib2026}). The resulting acceleration terms are (\acca,  \accd)=($-0.01\pm0.03$,\,$-0.01\pm0.06$)~mas\,yr$^{-2}$, both consistent with zero within the uncertainties. Thus, from an astrometric perspective, the system can be treated as a single star with the currently available data, and our final fit to this source does not consider the constant acceleration components.

Two additional sources, V*\,V1230\,Ori and COUP\,450, are classified as suspected astrometric binaries, likely with long orbital periods. Both display significantly larger astrometric residuals after fitting linear motion and parallax compared to typical single stars. Including acceleration terms improves the fits, and significant detections are obtained. For V*\,V1230\,Ori, the accelerations (\acca,  \accd)=($-0.60\pm0.03$,$-1.83\pm0.25$)~mas\,yr$^{-2}$ are detected at the $20\sigma$ and $7\sigma$ levels in R.A. and Dec., respectively. For COUP\,450, we measure (\acca,  \accd)=($-0.14\pm0.04$,\,$0.08\pm0.05$)~mas\,yr$^{-2}$, with the R.A. component detected at a significance level of $3.5\sigma$. The implications of these acceleration detections are discussed further in Sect.~\ref{sec:discussion}.

The radio source associated with V* V1230 Ori is likely a companion to the massive B1 star seen at optical wavelengths. The radio-optical position difference of 35.80\,mas (see Table~6 in \citetalias{dzib2026}) supports this view. This will, a posteriori, explain the necessity of adding constant-acceleration terms to the motion fits.

\section{Discussion}\label{sec:discussion}

The VLBA astrometry presented in this work enables the first homogeneous dynamical characterization of multiple young stellar systems across the Orion complex. Beyond providing parallaxes and proper motions, the DYNAMO–VLBA survey allows the direct determination of stellar masses and orbital parameters, thereby offering crucial tests of PMS stellar evolution models and a benchmark for multi-wavelength studies of young binaries. The discussion below places our results in the context of these broader astrophysical questions.

We begin by comparing the derived dynamical masses with previous estimates obtained from spectroscopy, SED fitting, or earlier interferometric campaigns. These comparisons assess the internal consistency of our VLBA modeling and highlight cases in which discrepancies may reflect additional complexity, such as orbital inclination effects, limited temporal sampling, or unresolved multiplicity. For two well-characterized systems, Brun~656 and HD~294300, we further perform SED modeling to test the predictions of PMS evolutionary tracks against directly measured stellar masses.

For systems where only one radio component is detected, we discuss how the astrometric mass function and measured accelerations can still provide meaningful dynamical constraints on unseen companions. Finally, we examine the implications of our findings for the occurrence of magnetic activity in intermediate-mass young stars, particularly those exhibiting nonthermal radio emission, which may trace the onset of magnetism in a regime where standard convective dynamos are not expected to operate.

\subsection{Mass comparisons}

We have determined dynamical masses for seven systems (Table~\ref{tab:mass}) and compare them here with previously reported values listed in Table~\ref{tab:binaries}. This comparison provides a first benchmark for assessing the reliability of our results and identifying systems where discrepancies may indicate additional astrophysical complexity. Masses reported in the literature mainly were derived from spectroscopic orbital fitting or from placement in the Hertzsprung–Russell diagram and comparison with evolutionary models.  Only in the case of CXOU~J054146.1$-$015622 were the previously estimated masses also obtained from VLBA astrometry, albeit with fewer detections.

For several systems, such as Brun\,656, V*\,NU\,Ori, and HD\,294300, the VLBA-derived dynamical masses are broadly consistent with values obtained from spectroscopy or SED fitting, lending confidence to the robustness of our orbital modeling. In particular, Brun\,656 shows excellent agreement between our measurement ($2.64\pm0.13$\msun for the primary) and the SED-based estimate of $2.72^{+0.21}_{-0.15}$\msun by \citet{valegrad2021}, confirming its intermediate-mass nature. Likewise, for V*\,NU\,Ori, our dynamical mass estimate for component~C ($6.96\pm0.32$\msun) is consistent with the spectroscopic mass estimate of $7.8\pm0.7$\msun from \citet{schultz2019} for the same component, supporting its association of nonthermal radio emission and magnetic activity.

In contrast, for the close spectroscopic binaries Parenago\,1540 and HD\,37017, our dynamical masses are significantly lower than those previously reported. In the case of Parenago\,1540 our estimated masses are $M_1=0.34\pm0.04$\msun and M$_2=0.44\pm0.05$\msun, whereas earlier estimates were 1.7\msun and 1.25\msun, respectively. Similarly, for HD\,37017 we derive $M_1=2.28\pm0.07$\msun and M$_2=1.20\pm0.03$\msun, while previous estimates range from 3.5 to 8.5\msun for the main star and from 1.8 to 4.5\msun for its companion. 
This discrepancy most likely reflects the sparse temporal sampling of the available VLBA data (with cadences set by the GOBELINS survey) and the short orbital periods of the systems, which together limit the accuracy of orbital fitting. These results suggest that the formal uncertainties in these cases are underestimated, and that additional VLBA monitoring with improved cadence will be essential to obtain more reliable dynamical masses.

Among the four systems in which both stellar components were detected with the VLBA, the derived dynamical masses offer an exceptional opportunity for direct comparison with independent estimates from SED modeling. Such a comparison is particularly valuable because it allows the empirical calibration of PMS evolutionary tracks using dynamically measured stellar masses.  However, for two of these systems, CXOU~J054146.1$-$015622 and 2MASS~J05414134$-$0153326, the available photometric data are too limited to perform a reliable SED analysis. Therefore, in the following we focus on Brun~656 and HD~294300, the two systems with sufficient multiwavelength coverage to enable detailed SED modeling and comparison with PMS evolutionary predictions.

\begin{figure*}
    \centering
    \begin{subfigure}[b]{0.48\linewidth}
        \centering
        \includegraphics[width=\linewidth]{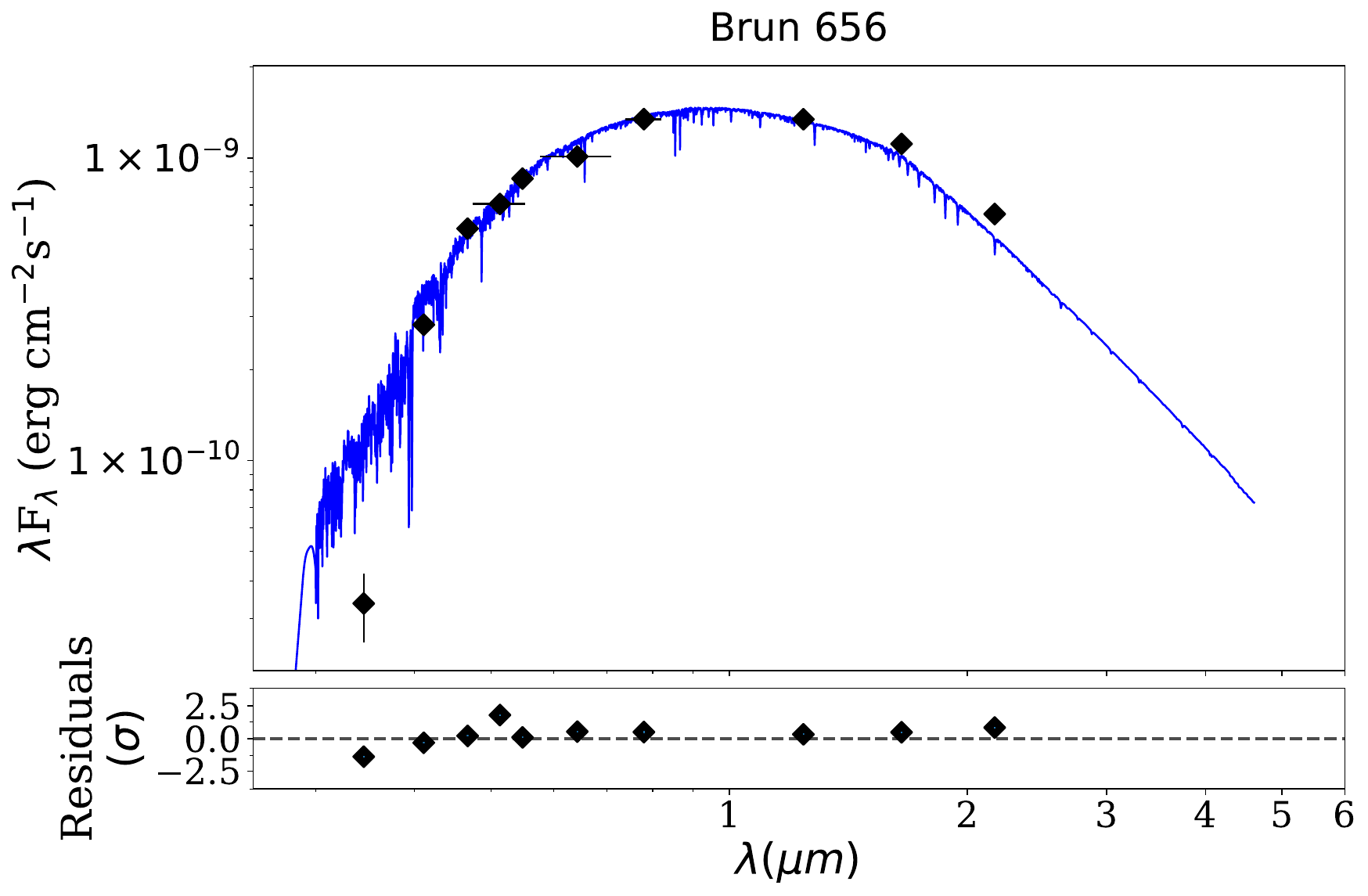}
        \label{fig:sed_brun}
    \end{subfigure}
    \hfill
    \begin{subfigure}[b]{0.48\linewidth}
        \centering
        \includegraphics[width=\linewidth]{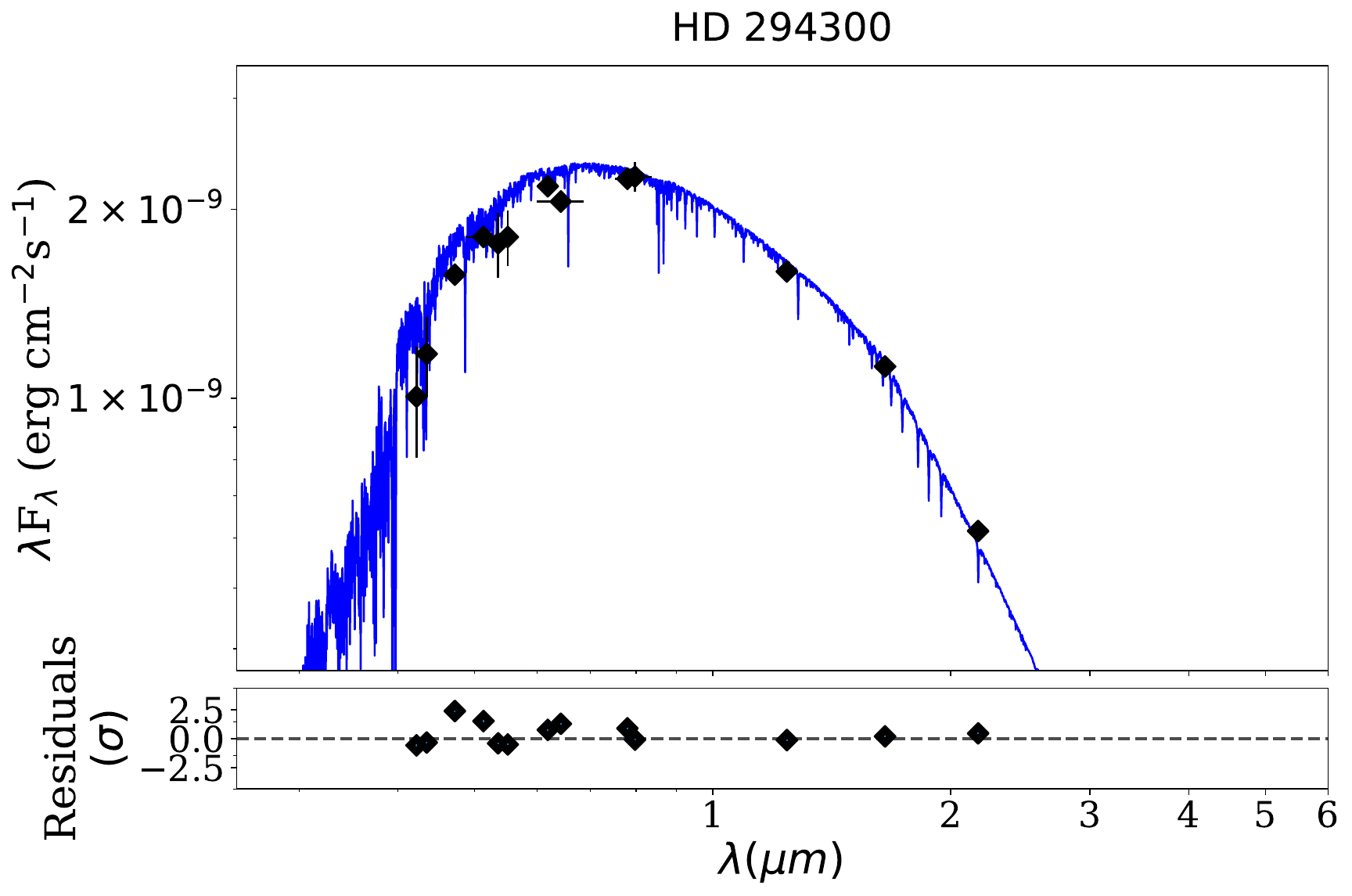}
        \label{fig:sed_hd294300}
    \end{subfigure}
    \caption{Spectral energy distributions of Brun~656 and HD~294300 obtained with \texttt{ARIADNE}. 
    The solid blue line represents the best-fit model, and black markers denote the observed photometry used in the fit.}
    \label{fig:sed}
\end{figure*}

\subsection{Spectral energy distribution (SED) modeling}
\label{sec:sed}

The precise determination of dynamical masses in binary systems provides a fundamental reference for calibrating PMS evolutionary models, which still show significant discrepancies, particularly in the intermediate–mass regime. A clear example is the previous study of the Oph~S1 system, where PMS models overpredicted the mass by 25–50\% relative to the dynamical value \citep{ordonez2024,ordonez2025a}. To extend such tests to the Orion sample, we analyze the SEDs of Brun~656 and HD~294300, the two binaries in which both components are resolved with the VLBA and have broad multi-wavelength photometric coverage.

Modeling the SEDs allows us to derive the stellar effective temperatures ($T_{\rm eff}$), bolometric luminosities ($L_\star$), and extinction values ($A_V$), thereby enabling a direct comparison between dynamical and model-predicted stellar masses. The photometric characterization and SED fitting were performed with the \texttt{ARIADNE} code \citep{VinesJenkins2022}, which implements a fully Bayesian approach with {Bayesian Model Averaging} \citep{Fragoso2018}, combining several stellar atmosphere libraries according to their Bayesian evidence. Multiband photometry was automatically compiled through \texttt{astroquery} \citep{Ginsburg2019} from \textit{Gaia}~DR3 \citep{Gaia2023}, 2MASS \citep{Skrutskie2006}, Str\"omgren-Paunzen \citep{Paunzen2015}, Str\"omgren–Hauck \citep{Hauck1998, Mermilliod1986}, TYCHO-2 \citep{Hog2000}, Johnson~UBV \citep{Mermilliod1986}, SDSS~DR12 \citep{Alam2015}, and TESS \citep{Stassun2019}; low-quality SkyMapper data were discarded. The SEDs were modeled using the \texttt{PHOENIX~v2} \citep{Husser2013}, \texttt{BT-Settl}, \texttt{BT-NextGen}, and \texttt{BT-Cond} \citep{Allard2011,Allard2012,Hauschildt1999}, and \texttt{Kurucz/CK04} \citep{Castelli2003} atmosphere grids. Extinction was treated as a free parameter using the law of \citet{Fitzpatrick1999} with $R_V=3.1$. 

Distances were fixed to $403\pm9$~pc for Brun~656 and $405\pm10$~pc for HD~294300 based on our VLBA parallaxes \citepalias{dzib2026}, and $T_{\rm eff}$ priors were set according to the literature spectral types \citep{walker1969,valegrad2021,pinzon2021}. Posterior sampling was performed with \texttt{dynesty} \citep{Speagle2019}. The resulting parameters are: for Brun~656, $L_\star=25.0^{+2.7}_{-2.2}\,L_\odot$ and $A_V=1.65\pm0.10$~mag; for HD~294300, $L_\star=25.7^{+2.3}_{-2.0}\,L_\odot$ and $A_V=0.67\pm0.06$~mag. In both systems, the SEDs are well reproduced from the optical to mid-infrared wavelengths without significant excess and residuals $<2.5\sigma$ (see Fig.~\ref{fig:sed}), indicating that the emission is dominated by stellar photospheres.

\begin{figure*}
    \centering
    \includegraphics[width=\textwidth]{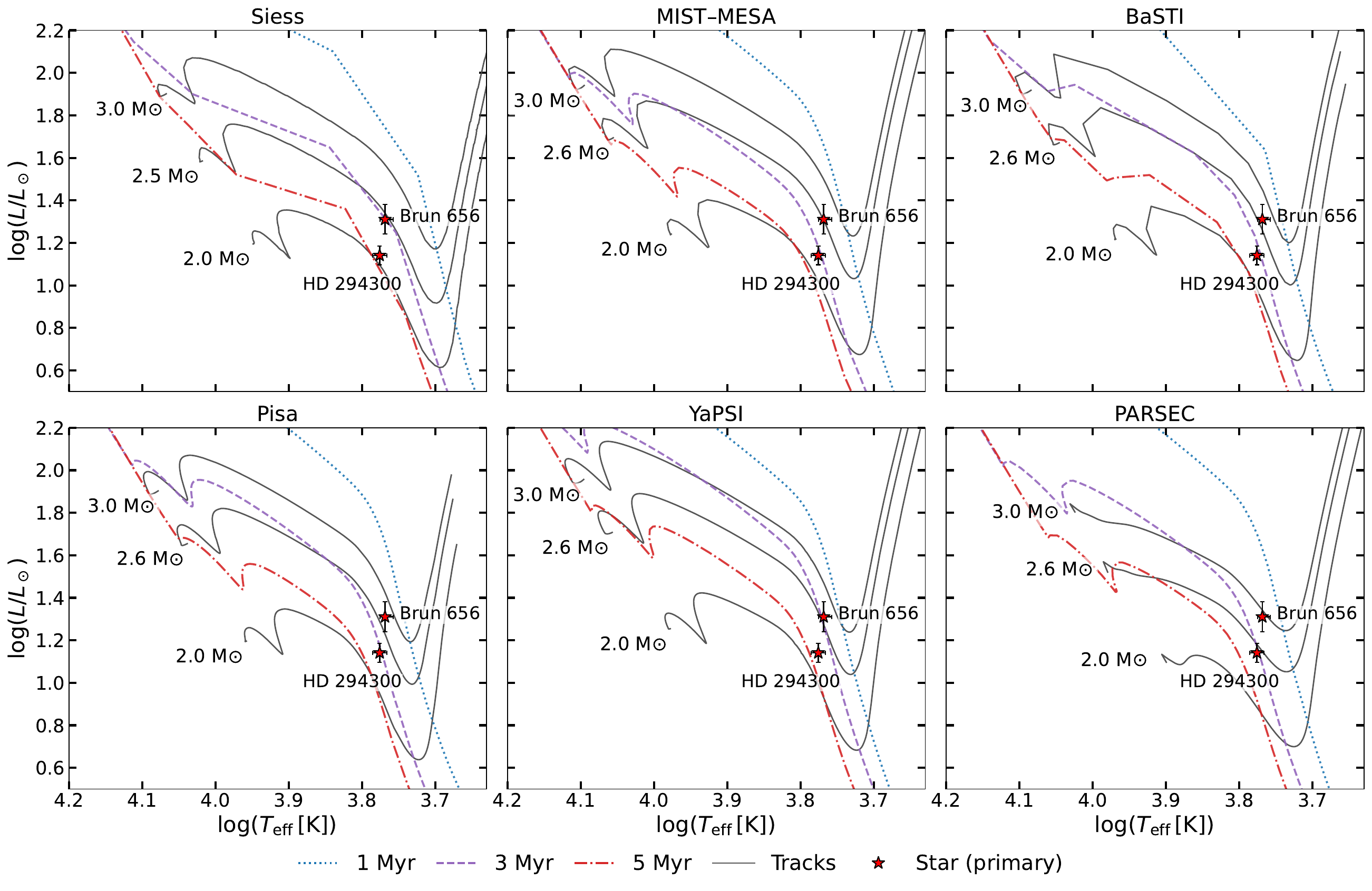}
    \caption{Location of the primary components of Brun~656 and HD~294300 on the HR diagram and evolutionary tracks based on pre–main-sequence stellar models.}
    \label{fig:models}
\end{figure*}

\subsection{Comparison with PMS evolutionary models}

Because the observed SED represents the combined emission of the system, it was necessary to estimate the individual contribution of each component to the total luminosity before comparing with PMS models. For each system, we used the total luminosity derived from the SED fit obtained with \texttt{ARIADNE}.  

To determine the contribution of each component, we employed PMS models from which theoretical luminosities $L_1$ and $L_2$ were extracted over an age range consistent with previous estimates: 1–3~Myr (Brun~656) and 1.5–3.5~Myr (HD~294300) \citep[e.g.,][]{valegrad2021,pinzon2021}. For each model, we computed the luminosity ratio between components and analyzed its evolution with age in order to quantify the relative contribution of each star to the system's total luminosity.  

Individual luminosities were estimated directly from the luminosity ratio predicted by the evolutionary models. For each evolutionary grid and within the adopted age range of each system, we computed the luminosities expected for the primary and secondary components at the dynamically determined masses and derived the corresponding luminosity ratio, $R_{L}=L_{1}/L_{2}$, on a fine age grid. The full set of $R_{L}$ values obtained by sampling all models and ages was then combined into a single distribution, which was summarized by its median, with percentile ranges used to quantify the associated dispersion. Assuming that the observed total luminosity satisfies $L_{\rm tot}=L_{1}+L_{2}$, the median luminosity ratio was applied to the total luminosity derived from the SED fit using \texttt{ARIADNE} to obtain the individual luminosities of each component. In this way, the 
luminosity partition reflects the theoretical proportions predicted by PMS models for the adopted dynamical masses and age ranges, 
while avoiding the introduction of multiple, model-dependent luminosity values.
 
The PMS models considered throughout this work (shown in Fig.~\ref{fig:models}) are: Siess \citep{Siess2000}, Pisa \citep{pisa2011}, MIST-MESA \citep{Choi2016, Paxton2011}, YaPSI \citep{Spada2017}, BaSTI \citep{Hidalgo2018} and  PARSEC \citep{Nguyen2022}.  

The results of the luminosity partition are: Brun~656 with $L_1=20.43^{+3.31}_{-3.26}\,L_\odot$ and $L_2=4.58^{+2.63}_{-2.65}\,L_\odot$; HD~294300 with $L_1=13.84^{+1.44}_{-1.40}\,L_\odot$ and $L_2=11.89^{+1.21}_{-1.25}\,L_\odot$. The effective temperatures used correspond to spectroscopic estimates representative of the optically dominant primary component. Figure~\ref{fig:models} shows the location of the primary components together with isochrones of 1, 3, and 5~Myr.

Anchoring the analysis to the VLBA dynamical masses ($M_{\rm Brun\,656}=2.64\pm0.13\,M_\odot$; $M_{\rm HD\,294300}=2.01\pm0.12\,M_\odot$), we derived the ages implied by each PMS grid. For Brun~656, all tracks place the star between the 2.5 and 3.0\,M$_\odot$ sequences, yielding ages of 2–3~Myr. For HD~294300, most models reproduce the observed position with a 2.0–2.5\,M$_\odot$ track at $\sim$3~Myr, except Siess and YaPSI, which require a slightly older age ($\sim$5~Myr). Overall, the scatter among models is modest ($\lesssim$0.3–0.5\,M$_\odot$ in mass or $\lesssim$2~Myr in age), and all are broadly consistent with the independent age estimates from optical studies.

Conversely, by adopting the literature age priors (1–3~Myr for Brun~656; 1.5–3.5~Myr for HD~294300) and reading off the corresponding model masses, a systematic difference emerges between the two systems.  
For Brun~656, the MIST–MESA, BaSTI, Pisa, and YaPSI grids reproduce the VLBA dynamical mass very well, predicting $M\simeq2.5$–$2.7\,M_\odot$, in close agreement with the measured value of $2.64\pm0.13\,M_\odot$.  
The Siess models yield slightly lower masses, while PARSEC predicts somewhat higher values ($\sim$3.0\,M$_\odot$).  
In contrast, for HD~294300 ($M_{\rm dyn}=2.01\pm0.12\,M_\odot$), nearly all models (MIST–MESA, BaSTI, Pisa, and PARSEC) predict systematically larger masses ($2.2$–$2.6\,M_\odot$) at the nominal 2–3~Myr age, whereas only the Siess grid yields a value consistent with the dynamical mass, but at an older age of $\sim$5~Myr.  
This behavior suggests that while current PMS models reproduce the evolutionary locus of Brun~656 remarkably well, they may overestimate luminosities (or, equivalently, underpredict contraction rates) for slightly less massive stars such as HD~294300.

Among the six grids explored, the MIST–MESA and BaSTI models achieve the best overall consistency between dynamical and theoretical predictions for both systems, followed closely by Pisa.  On the contrary, the PARSEC models show the larger disagreement by overpredicting the masses of both stars by $\sim0.5$\msun. 
The remaining models display small but systematic offsets of $\sim$0.2–0.4\,M$_\odot$, likely reflecting differences in input physics (e.g., convection efficiency, boundary conditions, or magnetic inhibition of convection). Overall, the DYNAMO–VLBA results highlight Brun~656 and HD~294300 as powerful empirical anchors for PMS model calibration in the intermediate–mass regime, providing direct, model-independent benchmarks to refine future evolutionary tracks.

\subsection{Masses of remaining stars}

While the dynamical masses of the remaining systems cannot be directly determined, the derived astrometric mass functions and acceleration vectors provide valuable constraints on the stellar masses and orbital configurations of these sources. Although the inferred parameters are subject to projection effects and limited time coverage, they nonetheless provide useful lower bounds on companion masses and orbital periods for embedded Orion systems where only one component is detected.

\subsubsection{The astrometric mass function}

The astrometric mass function, through Eq.~\ref{eq:3}, provides a relation between the companion mass, the primary mass, and the orbital inclination.
When the primary mass and inclination are constrained, the only remaining unknown in Eq.~\ref{eq:3} is the mass of the companion, $M_2$. By numerically solving Eq.~\ref{eq:3}, the mass function can therefore be used to estimate the mass of the companion. 

From our orbital fitting, we derived the astrometric mass function $f(M)$ (Table~\ref{tab:mass}) and the orbital inclinations (Table~\ref{tab:dm}) for four systems. Two of these systems, V*~MT~Ori and V*~V1399~Ori, have known spectral types (SpTs) from the literature (see Table~\ref{tab:binaries}), which allow us to estimate the masses of their primary components independently.  Their SpTs were converted to effective temperatures ($T_{\rm eff}$) using the intrinsic scale for young stars from \citet{Pecaut2013} and assuming an age of 3~Myr for both sources, consistent with typical ages of ONC members \citep[e.g.,][]{DaRio2010,Kounkel2018}\footnote{The $T_{\rm eff}$ calibration of \citet{Pecaut2013} applies to stars aged 5–30~Myr, but comparable intrinsic temperatures are expected for slightly younger PMS stars.}.  

The variable star MT~Ori (SpT$\sim$K2) corresponds to a $1.0\pm0.1$\msun young star, while V*~V1399~Ori (SpT$\sim$G8) corresponds to a $1.3\pm0.1$\msun star. Using these primary masses, together with orbital inclinations listed in Table~\ref{tab:dm}, and the corresponding values of $f(M)$, we solve Eq.\,\ref{eq:3}  to infer companion masses.  The resulting values are $0.27^{+2.88}_{-0.14}$\msun and $0.63^{+0.07}_{-0.06}$\msun for MT~Ori and V1399~Ori, respectively.

\subsubsection{Acceleration}

Constant angular acceleration parameters were derived for two additional stars, COUP~450 (GMR~A) and V*~V1230~Ori. The measured angular accelerations can be used to constrain the lower limit of the mass of the radio source companion if the projected separation is known. 
Moreover, the acceleration vector is expected to point toward the companion.

COUP~450 is a deeply embedded object previously detected only in radio and X-rays.  As discussed by \citet{dzib2021}, its measured proper motion varies between VLBA epochs, and our results confirm this behavior. However, because no secondary component has been detected or reported, further characterization of the system is not possible at this stage.

V*~V1230~Ori, on the other hand, is associated with a B1-type star. As noted by \citetalias{dzib2026}, the separation between the VLBA radio source and the position reported in \textit{Gaia}~DR3 is 35.8~mas, indicating that the radio emission likely arises from a lower-mass companion rather than from the massive B1 star itself.  As shown in Fig.~\ref{fig:Obi1}, the measured acceleration vector points toward the optical primary, as expected if it dominates the gravitational potential of the system.  In this context, the derived angular acceleration of the radio source can be used to estimate a lower limit for the mass of the massive optical star.

For simplicity, and in the absence of additional orbital information, we assume a circular and face-on orbit.  From Newton's law,
\[
m\,\mathtt{a}_{\rm tot} = \frac{G\,m\,M}{r^2} \;\Rightarrow\; M = \frac{\mathtt{a}_{\rm tot}\,r^2}{G},
\]
\noindent where $r$ is the projected separation between the stars, $\mathtt{a}_{\rm tot}$ is the physical acceleration of the radio source, and $M$ and $m$ denote the masses of the primary and secondary components, respectively.  We adopt $r \simeq 14$~AU ($2.1\times10^{14}$~cm) and a total angular acceleration of $\dot{\mu}_{\rm tot}=1.9$\mpys which corresponds to physical acceleration of $\mathtt{a}_{\rm tot}=0.011$~cm\,s$^{-2}$. These values yield a minimum mass of $M\simeq3.7$\msun for the optical star, consistent with its identification as a B1-type star, whose expected mass is $\sim12$\msun.

Assuming the radio-emitting component is less massive than the optical star, we can also estimate a rough orbital period. Taking a total system mass between 12 and 20~M$_\odot$ and the observed projected separation ($\sim$14~AU) as the semi-major axis, we obtain a period of $P\simeq12$–15~yr.  We note, however, that this represents only a lower limit, as the true period depends on the (unknown) orbital inclination and eccentricity.

\subsection{Intermediate mass stars with nonthermal radio emission}

In VLBI observations, the detection of radio emission at milliarcsecond scales provides strong evidence for a nonthermal emission mechanism, as it implies sources with brightness temperatures T$_{\rm b}>10^6$\,K, well above those attainable by thermal free--free emission \citep{thompson2017}. While not a primary objective of the DYNAMO-VLBA project, the sensitivity and angular resolution of our observations allow us to identify nonthermal emission associated with some intermediate-mass systems.

Nonthermal radio emission is commonly interpreted as a signature of magnetic activity. However, magnetic activity is generally not expected in intermediate- and high-mass stars, since their interiors are dominated by radiative energy transport and they lack the convective envelopes required to sustain a solar-type dynamo \citep{kraft1967,Mestel1968}. In particular, stars more massive than $\sim$2\,M$_\odot$ are predicted to evolve with radiative outer layers throughout most of their PMS and main-sequence lifetimes, preventing the magnetic field amplification mechanisms observed in lower-mass T~Tauri stars \citep{Gregory2012}. Nevertheless, several observational studies have revealed signatures of strong magnetic fields and activity in some young intermediate-mass stars. An initial explanation was that these signatures arise not from the intermediate-mass star itself, but from an unresolved lower-mass companion. Observationally, in the ONC, hundreds of X-ray–emitting A- and late-B–type stars have been detected whose temporal variability (e.g., flare-like light curves and timescales) is statistically indistinguishable from that of magnetically active T Tauri stars, strongly suggesting the contribution of hidden low-mass companions \citep{Montmerle2005}. Similarly, studies of Herbig Ae/Be systems indicate that X-ray emission is often due to chance or physical companions rather than intrinsic stellar activity \citep{Stelzer2006}. However, recent results have found evidence of magnetic activity directly associated with young intermediate-mass stars. These are often attributed either to fossil magnetic fields retained from the star formation process \citep{Donati2009,Bagnulo2020}, or to magnetic interactions within close binary systems where tidal forces or accretion may trigger dynamo-like activity \citep{Stelzer2009,Alecian2013}.  The detection of nonthermal radio emission in intermediate-mass YSOs thus provides critical constraints on the origin and evolution of magnetism in this transitional stellar mass regime.

Using the dynamical masses derived in this work together with the compact VLBI detections, we identified the following intermediate-mass stars associated with nonthermal radio emission: Brun\,656, HD 294300, and V* NU\,Ori\,C. In the case of HD\,37017, our determined dynamical mass is lower than spectroscopic estimates, likely due to underestimated errors (see Sect.~\ref{sec:results}). However, \citet{hohle2010} pointed out that the luminosity of this system is consistent with a star with a mass of 8.5\msun and, consequently, a companion of 4.5\msun. As the radio source is associated with both or any of these stars, our result is, in any case, consistent with nonthermal radio emission arising from an intermediate-mass star. 

Additionally, other observed radio sources fall into the category of intermediate-mass stars with nonthermal radio emission. Although no dynamical masses were estimated for them in this work, their classification is supported by their spectral type and by the small angular separation between our radio and \gaia results. These stars are HD\,290862, [SSC75]\,M78~11, $\theta_1$\,Ori\,E, HD\,37150, TYC\,5346-538-1, HD 294264, and V* KO Ori.  Finally, $\theta_1$\,Ori\,A$_2$ is a 4\msun star, and its radio emission is well known to be nonthermal, and it also falls within the present category. 

Altogether, intermediate-mass stars with nonthermal radio emission are 12 of the 54 stellar systems with VLBA fitted astrometry \citep[see also][]{dzib2026}, corresponding to $\sim22$\% of our sample. This fraction suggests that radio emission from magnetic activity is relatively common in this mass regime.

Of particular interest is the radio source associated with star {V*\,NU\,Ori\,C}. The derived dynamical mass $M_{\rm C,\,VLBA}=6.96\pm0.32$\msun is in good agreement with the result of $M_{\rm C}=7.8\pm0.7$\msun by \citet{schultz2019}.  This directly associates a nonthermal radio source with an intermediate-mass star lying just below the high-mass-star limit (M\,$\geq8$\msun). Previously, it was suggested that Oph\,S1 could represent another such case, as it was commonly assumed to have a mass between 6 and 8\,\msun. However, recent work by \citet{ordonez2024,ordonez2025a} has shown that Oph\,S1 instead has a mass of $4.12\pm0.04$\msun. Since the other known nonthermal sources discussed above are consistent with spectral types corresponding to masses $<5$\msun, V*\,NU\,Ori\,C remains a unique example of a nonthermal radio emitter with a mass close to the high-mass-star boundary.

Because the radio sources in our sample are all compact, with sizes of only a few mas, their emission is most plausibly linked to magnetic activity, as suggested by previous X-ray studies. A detailed exploration of the physical origin of this activity is beyond the scope of the present paper, but it clearly requires a coordinated multi-wavelength approach to fully disentangle the mechanisms at work.

\section{Conclusions} \label{sec:conclusions}

We have presented the results of the DYNAMO–VLBA survey focused on young binary and multiple systems within the Orion complex. In our multi–epoch VLBA observations,  we resolved multiple radio components in the HD\,294300 and 2MASS~J05414134$-$0153326 systems, confirming their binary nature. More generally, we obtained precise astrometric measurements that allowed us to resolve or detect orbital motions in 15 systems and to derive dynamical masses for seven of them. These results represent one of the most comprehensive VLBI determinations of stellar masses for young binaries in a single star–forming complex. 

For four systems in which both components are detected at radio wavelengths, the derived orbital solutions yield fully model-independent dynamical masses.  In particular, Brun~656 and HD~294300 exhibit excellent agreement between their VLBA–based masses and values inferred from spectral energy distribution modeling, supporting the reliability of our orbital fits and providing valuable empirical benchmarks for pre–main–sequence evolutionary models. For the spectroscopic binaries V*~NU~Ori, Parenago~1540, and HD~37017, our combination of VLBA astrometry and published spectroscopic constraints provides updated dynamical estimates, although the limited temporal sampling of the VLBA data leads to increased uncertainties for the shortest–period systems.

In addition to these binaries with complete orbits, we identified several systems exhibiting clear astrometric signatures of multiplicity, as evidenced by acceleration or periodic residuals.  For these, we derived astrometric mass functions or lower limits to companion masses, extending dynamical constraints to systems in which only one component is detected.  The detection of significant accelerations in sources such as COUP~450 and V*~V1230~Ori further demonstrates the capability of VLBI to uncover companions at separations of a few astronomical units, even in embedded or optically obscured regions.

Our results also reveal that a substantial fraction of the radio–emitting young stars in Orion are intermediate-mass objects exhibiting nonthermal emission.  Systems such as V*~NU~Ori~C, Brun~656, and HD~294300 provide direct evidence of magnetic activity in stars near or above the limit where standard convective dynamos are expected to cease, offering key observational constraints on the origin of magnetism in this transitional mass regime.

Altogether, the DYNAMO–VLBA observations establish VLBI astrometry as a uniquely powerful tool for measuring stellar masses and orbital dynamics in young multiple systems.  The derived dynamical masses, combined with SED–based luminosities and future spectroscopic monitoring, will provide essential benchmarks for refining PMS evolutionary models and for advancing our understanding of magnetic activity and multiplicity among young intermediate-mass stars.

\begin{acknowledgements}
We thank the anonymous referee for a careful and constructive review that helped improve the clarity and presentation of this work. 
S.A.D. acknowledges the M2FINDERS project from the European Research
Council (ERC) under the European Union's Horizon 2020 research and innovation programme
(grant No 101018682).
G.N.O.L. acknowledges the financial support provided by Secretaría de Ciencia, Humanidades, Tecnología e Innovación (Secihti) through grant CBF-2025-I-201.
P.A.B.G. acknowledges financial support from the São Paulo Research Foundation (FAPESP, grant: 2020/12518-8) and Conselho Nacional de Desenvolvimento Científico e Tecnológico (CNPq, grant: 303659/2024-6).
J.M.M.S. acknowledges financial support from the State Agency for Research of the Spanish Ministry of Science and Innovation under grants PID2022-136828NB-C41 and EX2024-001451-M funded by MICIU/AEI/10.13039/501100011033/ERDF/EU.
The National Radio Astronomy Observatory is a facility of the National Science Foundation operated under cooperative agreement by Associated Universities, Inc.
\end{acknowledgements}

\bibliographystyle{aa_url} 
\bibliography{references.bib} 

@article{Fragoso2018,
author = {Fragoso, Tiago M. and Bertoli, Wesley and Louzada, Francisco},
title = {Bayesian Model Averaging: A Systematic Review and Conceptual Classification},
journal = {International Statistical Review},
volume = {86},
number = {1},
pages = {1-28},
doi = {https://doi.org/10.1111/insr.12243},
url = {https://onlinelibrary.wiley.com/doi/abs/10.1111/insr.12243},
eprint = {https://onlinelibrary.wiley.com/doi/pdf/10.1111/insr.12243},
year = {2018}
}

@ARTICLE{jeffries2007,
       author = {{Jeffries}, R.~D.},
        title = "{The distance to the Orion Nebula cluster}",
      journal = {\mnras},
         year = 2007,
        month = apr,
       volume = {376},
       number = {3},
        pages = {1109-1119},
          doi = {10.1111/j.1365-2966.2007.11471.x},
archivePrefix = {arXiv},
       eprint = {astro-ph/0701186},
 primaryClass = {astro-ph},
       adsurl = {https://ui.adsabs.harvard.edu/abs/2007MNRAS.376.1109J}
}

@ARTICLE{zhang2020,
       author = {{Zhang}, Qicheng and {Hallinan}, Gregg and {Brisken}, Walter and {Bourke}, Stephen and {Golden}, Aaron},
        title = "{Multiepoch VLBI of L Dwarf Binary 2MASS J0746+2000AB: Precise Mass Measurements and Confirmation of Radio Emission from Both Components}",
      journal = {\apj},
         year = 2020,
        month = jul,
       volume = {897},
       number = {1},
          eid = {11},
        pages = {11},
          doi = {10.3847/1538-4357/ab9177},
archivePrefix = {arXiv},
       eprint = {2005.03657},
 primaryClass = {astro-ph.SR},
       adsurl = {https://ui.adsabs.harvard.edu/abs/2020ApJ...897...11Z}
}

@ARTICLE{azulay2017,
       author = {{Azulay}, R. and {Guirado}, J.~C. and {Marcaide}, J.~M. and {Mart{\'\i}-Vidal}, I. and {Ros}, E. and {Tognelli}, E. and {Jauncey}, D.~L. and {Lestrade}, J.-F. and {Reynolds}, J.~E.},
        title = "{The AB Doradus system revisited: The dynamical mass of AB Dor A/C}",
      journal = {\aap},
         year = 2017,
        month = oct,
       volume = {607},
          eid = {A10},
        pages = {A10},
          doi = {10.1051/0004-6361/201730641},
       adsurl = {https://ui.adsabs.harvard.edu/abs/2017A&A...607A..10A}
}

@ARTICLE{Alam2015,
       author = {{Alam}, Shadab and {Albareti}, Franco D. and {Allende Prieto}, Carlos and {Anders}, F. and {Anderson}, Scott F. and {Anderton}, Timothy and {Andrews}, Brett H. and {Armengaud}, Eric and {Aubourg}, {\'E}ric and {Bailey}, Stephen and {Basu}, Sarbani and {Bautista}, Julian E. and {Beaton}, Rachael L. and {Beers}, Timothy C. and {Bender}, Chad F. and {Berlind}, Andreas A. and {Beutler}, Florian and {Bhardwaj}, Vaishali and {Bird}, Jonathan C. and {Bizyaev}, Dmitry and {Blake}, Cullen H. and {Blanton}, Michael R. and {Blomqvist}, Michael and {Bochanski}, John J. and {Bolton}, Adam S. and {Bovy}, Jo and {Shelden Bradley}, A. and {Brandt}, W.~N. and {Brauer}, D.~E. and {Brinkmann}, J. and {Brown}, Peter J. and {Brownstein}, Joel R. and {Burden}, Angela and {Burtin}, Etienne and {Busca}, Nicol{\'a}s G. and {Cai}, Zheng and {Capozzi}, Diego and {Carnero Rosell}, Aurelio and {Carr}, Michael A. and {Carrera}, Ricardo and {Chambers}, K.~C. and {Chaplin}, William James and {Chen}, Yen-Chi and {Chiappini}, Cristina and {Chojnowski}, S. Drew and {Chuang}, Chia-Hsun and {Clerc}, Nicolas and {Comparat}, Johan and {Covey}, Kevin and {Croft}, Rupert A.~C. and {Cuesta}, Antonio J. and {Cunha}, Katia and {da Costa}, Luiz N. and {Da Rio}, Nicola and {Davenport}, James R.~A. and {Dawson}, Kyle S. and {De Lee}, Nathan and {Delubac}, Timoth{\'e}e and {Deshpande}, Rohit and {Dhital}, Saurav and {Dutra-Ferreira}, Let{\'\i}cia and {Dwelly}, Tom and {Ealet}, Anne and {Ebelke}, Garrett L. and {Edmondson}, Edward M. and {Eisenstein}, Daniel J. and {Ellsworth}, Tristan and {Elsworth}, Yvonne and {Epstein}, Courtney R. and {Eracleous}, Michael and {Escoffier}, Stephanie and {Esposito}, Massimiliano and {Evans}, Michael L. and {Fan}, Xiaohui and {Fern{\'a}ndez-Alvar}, Emma and {Feuillet}, Diane and {Filiz Ak}, Nurten and {Finley}, Hayley and {Finoguenov}, Alexis and {Flaherty}, Kevin and {Fleming}, Scott W. and {Font-Ribera}, Andreu and {Foster}, Jonathan and {Frinchaboy}, Peter M. and {Galbraith-Frew}, J.~G. and {Garc{\'\i}a}, Rafael A. and {Garc{\'\i}a-Hern{\'a}ndez}, D.~A. and {Garc{\'\i}a P{\'e}rez}, Ana E. and {Gaulme}, Patrick and {Ge}, Jian and {G{\'e}nova-Santos}, R. and {Georgakakis}, A. and {Ghezzi}, Luan and {Gillespie}, Bruce A. and {Girardi}, L{\'e}o and {Goddard}, Daniel and {Gontcho}, Satya Gontcho A. and {Gonz{\'a}lez Hern{\'a}ndez}, Jonay I. and {Grebel}, Eva K. and {Green}, Paul J. and {Grieb}, Jan Niklas and {Grieves}, Nolan and {Gunn}, James E. and {Guo}, Hong and {Harding}, Paul and {Hasselquist}, Sten and {Hawley}, Suzanne L. and {Hayden}, Michael and {Hearty}, Fred R. and {Hekker}, Saskia and {Ho}, Shirley and {Hogg}, David W. and {Holley-Bockelmann}, Kelly and {Holtzman}, Jon A. and {Honscheid}, Klaus and {Huber}, Daniel and {Huehnerhoff}, Joseph and {Ivans}, Inese I. and {Jiang}, Linhua and {Johnson}, Jennifer A. and {Kinemuchi}, Karen and {Kirkby}, David and {Kitaura}, Francisco and {Klaene}, Mark A. and {Knapp}, Gillian R. and {Kneib}, Jean-Paul and {Koenig}, Xavier P. and {Lam}, Charles R. and {Lan}, Ting-Wen and {Lang}, Dustin and {Laurent}, Pierre and {Le Goff}, Jean-Marc and {Leauthaud}, Alexie and {Lee}, Khee-Gan and {Lee}, Young Sun and {Licquia}, Timothy C. and {Liu}, Jian and {Long}, Daniel C. and {L{\'o}pez-Corredoira}, Mart{\'\i}n and {Lorenzo-Oliveira}, Diego and {Lucatello}, Sara and {Lundgren}, Britt and {Lupton}, Robert H. and {Mack}, III, Claude E. and {Mahadevan}, Suvrath and {Maia}, Marcio A.~G. and {Majewski}, Steven R. and {Malanushenko}, Elena and {Malanushenko}, Viktor and {Manchado}, A. and {Manera}, Marc and {Mao}, Qingqing and {Maraston}, Claudia and {Marchwinski}, Robert C. and {Margala}, Daniel and {Martell}, Sarah L. and {Martig}, Marie and {Masters}, Karen L. and {Mathur}, Savita and {McBride}, Cameron K. and {McGehee}, Peregrine M. and {McGreer}, Ian D. and {McMahon}, Richard G. and {M{\'e}nard}, Brice and {Menzel}, Marie-Luise and {Merloni}, Andrea and {M{\'e}sz{\'a}ros}, Szabolcs and {Miller}, Adam A. and {Miralda-Escud{\'e}}, Jordi and {Miyatake}, Hironao and {Montero-Dorta}, Antonio D. and {More}, Surhud and {Morganson}, Eric and {Morice-Atkinson}, Xan and {Morrison}, Heather L. and {Mosser}, Ben{\^o}it and {Muna}, Demitri and {Myers}, Adam D. and {Nandra}, Kirpal and {Newman}, Jeffrey A. and {Neyrinck}, Mark and {Nguyen}, Duy Cuong and {Nichol}, Robert C. and {Nidever}, David L. and {Noterdaeme}, Pasquier and {Nuza}, Sebasti{\'a}n E. and {O'Connell}, Julia E. and {O'Connell}, Robert W. and {O'Connell}, Ross and {Ogando}, Ricardo L.~C. and {Olmstead}, Matthew D. and {Oravetz}, Audrey E. and {Oravetz}, Daniel J. and {Osumi}, Keisuke and {Owen}, Russell and {Padgett}, Deborah L. and {Padmanabhan}, Nikhil and {Paegert}, Martin and {Palanque-Delabrouille}, Nathalie and {Pan}, Kaike},
        title = "{The Eleventh and Twelfth Data Releases of the Sloan Digital Sky Survey: Final Data from SDSS-III}",
      journal = {\apjs},
         year = 2015,
        month = jul,
       volume = {219},
       number = {1},
          eid = {12},
        pages = {12},
          doi = {10.1088/0067-0049/219/1/12},
archivePrefix = {arXiv},
       eprint = {1501.00963},
 primaryClass = {astro-ph.IM},
       adsurl = {https://ui.adsabs.harvard.edu/abs/2015ApJS..219...12A}
}

@ARTICLE{Mermilliod1986,
       author = {{Mermilliod}, J.-C.},
        title = "{Compilation of Eggen's UBV data, transformed to UBV (unpublished)}",
      journal = {Catalogue of Eggen's UBV data},
         year = 1986,
        month = jan,
        pages = {0},
       adsurl = {https://ui.adsabs.harvard.edu/abs/1986EgUBV........0M}
}

@ARTICLE{Hog2000,
       author = {{H{\o}g}, E. and {Fabricius}, C. and {Makarov}, V.~V. and {Urban}, S. and {Corbin}, T. and {Wycoff}, G. and {Bastian}, U. and {Schwekendiek}, P. and {Wicenec}, A.},
        title = "{The Tycho-2 catalogue of the 2.5 million brightest stars}",
      journal = {\aap},
         year = 2000,
        month = mar,
       volume = {355},
        pages = {L27-L30},
       adsurl = {https://ui.adsabs.harvard.edu/abs/2000A&A...355L..27H}
}

@ARTICLE{Hauck1998,
       author = {{Hauck}, B. and {Mermilliod}, M.},
        title = "{Uvbybeta photoelectric photometric catalogue}",
      journal = {\aaps},
         year = 1998,
        month = may,
       volume = {129},
        pages = {431-433},
          doi = {10.1051/aas:1998195},
       adsurl = {https://ui.adsabs.harvard.edu/abs/1998A&AS..129..431H}
}

@ARTICLE{Paunzen2015,
       author = {{Paunzen}, E.},
        title = "{A new catalogue of Str{\"o}mgren-Crawford uvby{\ensuremath{\beta}} photometry}",
      journal = {\aap},
         year = 2015,
        month = aug,
       volume = {580},
          eid = {A23},
        pages = {A23},
          doi = {10.1051/0004-6361/201526413},
archivePrefix = {arXiv},
       eprint = {1506.04568},
 primaryClass = {astro-ph.SR},
       adsurl = {https://ui.adsabs.harvard.edu/abs/2015A&A...580A..23P}
}

@ARTICLE{Stassun2019,
       author = {{Stassun}, Keivan G. and {Oelkers}, Ryan J. and {Paegert}, Martin and {Torres}, Guillermo and {Pepper}, Joshua and {De Lee}, Nathan and {Collins}, Kevin and {Latham}, David W. and {Muirhead}, Philip S. and {Chittidi}, Jay and {Rojas-Ayala}, B{\'a}rbara and {Fleming}, Scott W. and {Rose}, Mark E. and {Tenenbaum}, Peter and {Ting}, Eric B. and {Kane}, Stephen R. and {Barclay}, Thomas and {Bean}, Jacob L. and {Brassuer}, C.~E. and {Charbonneau}, David and {Ge}, Jian and {Lissauer}, Jack J. and {Mann}, Andrew W. and {McLean}, Brian and {Mullally}, Susan and {Narita}, Norio and {Plavchan}, Peter and {Ricker}, George R. and {Sasselov}, Dimitar and {Seager}, S. and {Sharma}, Sanjib and {Shiao}, Bernie and {Sozzetti}, Alessandro and {Stello}, Dennis and {Vanderspek}, Roland and {Wallace}, Geoff and {Winn}, Joshua N.},
        title = "{The Revised TESS Input Catalog and Candidate Target List}",
      journal = {\aj},
         year = 2019,
        month = oct,
       volume = {158},
       number = {4},
          eid = {138},
        pages = {138},
          doi = {10.3847/1538-3881/ab3467},
archivePrefix = {arXiv},
       eprint = {1905.10694},
 primaryClass = {astro-ph.SR},
       adsurl = {https://ui.adsabs.harvard.edu/abs/2019AJ....158..138S}
}

@ARTICLE{Skrutskie2006,
       author = {{Skrutskie}, M.~F. and {Cutri}, R.~M. and {Stiening}, R. and {Weinberg}, M.~D. and {Schneider}, S. and {Carpenter}, J.~M. and {Beichman}, C. and {Capps}, R. and {Chester}, T. and {Elias}, J. and {Huchra}, J. and {Liebert}, J. and {Lonsdale}, C. and {Monet}, D.~G. and {Price}, S. and {Seitzer}, P. and {Jarrett}, T. and {Kirkpatrick}, J.~D. and {Gizis}, J.~E. and {Howard}, E. and {Evans}, T. and {Fowler}, J. and {Fullmer}, L. and {Hurt}, R. and {Light}, R. and {Kopan}, E.~L. and {Marsh}, K.~A. and {McCallon}, H.~L. and {Tam}, R. and {Van Dyk}, S. and {Wheelock}, S.},
        title = "{The Two Micron All Sky Survey (2MASS)}",
      journal = {\aj},
         year = 2006,
        month = feb,
       volume = {131},
       number = {2},
        pages = {1163-1183},
          doi = {10.1086/498708},
       adsurl = {https://ui.adsabs.harvard.edu/abs/2006AJ....131.1163S}
}

@ARTICLE{Hidalgo2018,
       author = {{Hidalgo}, Sebastian L. and {Pietrinferni}, Adriano and {Cassisi}, Santi and {Salaris}, Maurizio and {Mucciarelli}, Alessio and {Savino}, Alessandro and {Aparicio}, Antonio and {Silva Aguirre}, Victor and {Verma}, Kuldeep},
        title = "{The Updated BaSTI Stellar Evolution Models and Isochrones. I. Solar-scaled Calculations}",
      journal = {\apj},
         year = 2018,
        month = apr,
       volume = {856},
       number = {2},
          eid = {125},
        pages = {125},
          doi = {10.3847/1538-4357/aab158},
archivePrefix = {arXiv},
       eprint = {1802.07319},
 primaryClass = {astro-ph.GA},
       adsurl = {https://ui.adsabs.harvard.edu/abs/2018ApJ...856..125H}
}

@ARTICLE{Nguyen2022,
       author = {{Nguyen}, C.~T. and {Costa}, G. and {Girardi}, L. and {Volpato}, G. and {Bressan}, A. and {Chen}, Y. and {Marigo}, P. and {Fu}, X. and {Goudfrooij}, P.},
        title = "{PARSEC V2.0: Stellar tracks and isochrones of low- and intermediate-mass stars with rotation}",
      journal = {\aap},
         year = 2022,
        month = sep,
       volume = {665},
          eid = {A126},
        pages = {A126},
          doi = {10.1051/0004-6361/202244166},
archivePrefix = {arXiv},
       eprint = {2207.08642},
 primaryClass = {astro-ph.SR},
       adsurl = {https://ui.adsabs.harvard.edu/abs/2022A&A...665A.126N}
}

@ARTICLE{Spada2017,
       author = {{Spada}, F. and {Demarque}, P. and {Kim}, Y.-C. and {Boyajian}, T.~S. and {Brewer}, J.~M.},
        title = "{The Yale-Potsdam Stellar Isochrones}",
      journal = {\apj},
         year = 2017,
        month = apr,
       volume = {838},
       number = {2},
          eid = {161},
        pages = {161},
          doi = {10.3847/1538-4357/aa661d},
archivePrefix = {arXiv},
       eprint = {1703.03975},
 primaryClass = {astro-ph.SR},
       adsurl = {https://ui.adsabs.harvard.edu/abs/2017ApJ...838..161S}
}

@ARTICLE{Paxton2011,
       author = {{Paxton}, Bill and {Bildsten}, Lars and {Dotter}, Aaron and {Herwig}, Falk and {Lesaffre}, Pierre and {Timmes}, Frank},
        title = "{Modules for Experiments in Stellar Astrophysics (MESA)}",
      journal = {\apjs},
         year = 2011,
        month = jan,
       volume = {192},
       number = {1},
          eid = {3},
        pages = {3},
          doi = {10.1088/0067-0049/192/1/3},
archivePrefix = {arXiv},
       eprint = {1009.1622},
 primaryClass = {astro-ph.SR},
       adsurl = {https://ui.adsabs.harvard.edu/abs/2011ApJS..192....3P}
}

@ARTICLE{Choi2016,
       author = {{Choi}, Jieun and {Dotter}, Aaron and {Conroy}, Charlie and {Cantiello}, Matteo and {Paxton}, Bill and {Johnson}, Benjamin D.},
        title = "{Mesa Isochrones and Stellar Tracks (MIST). I. Solar-scaled Models}",
      journal = {\apj},
         year = 2016,
        month = jun,
       volume = {823},
       number = {2},
          eid = {102},
        pages = {102},
          doi = {10.3847/0004-637X/823/2/102},
archivePrefix = {arXiv},
       eprint = {1604.08592},
 primaryClass = {astro-ph.SR},
       adsurl = {https://ui.adsabs.harvard.edu/abs/2016ApJ...823..102C}
}

@INPROCEEDINGS{Castelli2003,
       author = {{Castelli}, F. and {Kurucz}, R.~L.},
        title = "{New Grids of ATLAS9 Model Atmospheres}",
    booktitle = {Modelling of Stellar Atmospheres},
         year = 2003,
       editor = {{Piskunov}, N. and {Weiss}, W.~W. and {Gray}, D.~F.},
       series = {IAU Symposium},
       volume = {210},
        month = jan,
        pages = {A20},
          doi = {10.48550/arXiv.astro-ph/0405087},
archivePrefix = {arXiv},
       eprint = {astro-ph/0405087},
 primaryClass = {astro-ph},
       adsurl = {https://ui.adsabs.harvard.edu/abs/2003IAUS..210P.A20C}
}

@ARTICLE{Allard2012,
       author = {{Allard}, F. and {Homeier}, D. and {Freytag}, B.},
        title = "{Models of very-low-mass stars, brown dwarfs and exoplanets}",
      journal = {Philosophical Transactions of the Royal Society of London Series A},
         year = 2012,
        month = jun,
       volume = {370},
       number = {1968},
        pages = {2765-2777},
          doi = {10.1098/rsta.2011.0269},
archivePrefix = {arXiv},
       eprint = {1112.3591},
 primaryClass = {astro-ph.SR},
       adsurl = {https://ui.adsabs.harvard.edu/abs/2012RSPTA.370.2765A}
}

@INPROCEEDINGS{Allard2011,
       author = {{Allard}, F. and {Homeier}, D. and {Freytag}, B.},
        title = "{Model Atmospheres From Very Low Mass Stars to Brown Dwarfs}",
    booktitle = {16th Cambridge Workshop on Cool Stars, Stellar Systems, and the Sun},
         year = 2011,
       editor = {{Johns-Krull}, Christopher and {Browning}, Matthew K. and {West}, Andrew A.},
       series = {Astronomical Society of the Pacific Conference Series},
       volume = {448},
        month = dec,
        pages = {91},
          doi = {10.48550/arXiv.1011.5405},
archivePrefix = {arXiv},
       eprint = {1011.5405},
 primaryClass = {astro-ph.SR},
       adsurl = {https://ui.adsabs.harvard.edu/abs/2011ASPC..448...91A}
}

@ARTICLE{Husser2013,
       author = {{Husser}, T.-O. and {Wende-von Berg}, S. and {Dreizler}, S. and {Homeier}, D. and {Reiners}, A. and {Barman}, T. and {Hauschildt}, P.~H.},
        title = "{A new extensive library of PHOENIX stellar atmospheres and synthetic spectra}",
      journal = {\aap},
         year = 2013,
        month = may,
       volume = {553},
          eid = {A6},
        pages = {A6},
          doi = {10.1051/0004-6361/201219058},
archivePrefix = {arXiv},
       eprint = {1303.5632},
 primaryClass = {astro-ph.SR},
       adsurl = {https://ui.adsabs.harvard.edu/abs/2013A&A...553A...6H}
}

@ARTICLE{Fitzpatrick1999,
       author = {{Fitzpatrick}, Edward L.},
        title = "{Correcting for the Effects of Interstellar Extinction}",
      journal = {\pasp},
         year = 1999,
        month = jan,
       volume = {111},
       number = {755},
        pages = {63-75},
          doi = {10.1086/316293},
archivePrefix = {arXiv},
       eprint = {astro-ph/9809387},
 primaryClass = {astro-ph},
       adsurl = {https://ui.adsabs.harvard.edu/abs/1999PASP..111...63F}
}

@ARTICLE{Ginsburg2019,
       author = {{Ginsburg}, Adam and {Sip{\H{o}}cz}, Brigitta M. and {Brasseur}, C.~E. and {Cowperthwaite}, Philip S. and {Craig}, Matthew W. and {Deil}, Christoph and {Guillochon}, James and {Guzman}, Giannina and {Liedtke}, Simon and {Lian Lim}, Pey and {Lockhart}, Kelly E. and {Mommert}, Michael and {Morris}, Brett M. and {Norman}, Henrik and {Parikh}, Madhura and {Persson}, Magnus V. and {Robitaille}, Thomas P. and {Segovia}, Juan-Carlos and {Singer}, Leo P. and {Tollerud}, Erik J. and {de Val-Borro}, Miguel and {Valtchanov}, Ivan and {Woillez}, Julien and {Astroquery Collaboration} and {a subset of astropy Collaboration}},
        title = "{astroquery: An Astronomical Web-querying Package in Python}",
      journal = {\aj},
         year = 2019,
        month = mar,
       volume = {157},
       number = {3},
          eid = {98},
        pages = {98},
          doi = {10.3847/1538-3881/aafc33},
archivePrefix = {arXiv},
       eprint = {1901.04520},
 primaryClass = {astro-ph.IM},
       adsurl = {https://ui.adsabs.harvard.edu/abs/2019AJ....157...98G}
}

@article{Hauschildt1999,
         author = {Hauschildt, Peter H and Allard, France and Baron, E},
         doi = {10.1086/430754},
        issn = {0004-637X},
     journal = {The Astrophysical Journal},
      number = {2},
       pages = {865--872},
       title = {{THE NEXTGEN MODEL ATMOSPHERE GRID FOR 3000 {\textless}= Teff {\textless}= 10,000 K}},
      volume = {629},
        year = {1999}
}

@ARTICLE{Speagle2019,
       author = {{Speagle}, Joshua S.},
        title = "{DYNESTY: a dynamic nested sampling package for estimating Bayesian posteriors and evidences}",
      journal = {\mnras},
         year = 2020,
        month = apr,
       volume = {493},
       number = {3},
        pages = {3132-3158},
          doi = {10.1093/mnras/staa278},
archivePrefix = {arXiv},
       eprint = {1904.02180},
 primaryClass = {astro-ph.IM},
       adsurl = {https://ui.adsabs.harvard.edu/abs/2020MNRAS.493.3132S}
}

@ARTICLE{VinesJenkins2022,
       author = {{Vines}, Jose I. and {Jenkins}, James S.},
        title = "{ARIADNE: measuring accurate and precise stellar parameters through SED fitting}",
      journal = {\mnras},
         year = 2022,
        month = jun,
       volume = {513},
       number = {2},
        pages = {2719-2731},
          doi = {10.1093/mnras/stac956},
archivePrefix = {arXiv},
       eprint = {2204.03769},
 primaryClass = {astro-ph.SR},
       adsurl = {https://ui.adsabs.harvard.edu/abs/2022MNRAS.513.2719V}
}

@ARTICLE{dzib2026,
       author = {{Dzib}, Sergio A. and {Ord{\'o}{\~n}ez-Toro}, Jazm{\'\i}n and {Loinard}, Laurent and {Kounkel}, Marina and {Ortiz-Leon}, Gisela and {Galli}, Phillip A.~B. and {Rodr{\'\i}guez}, Luis F. and {Mioduszewski}, Amy J. and {Masqu{\'e}}, Josep M. and {O'Kelly}, Eoin and {Forbrich}, Jan and {Moo-Herrera}, Karla},
        title = "{Dynamical masses of YSOs with the VLBA: DYNAMO VLBA: Trigonometric parallaxes and proper motions of YSOs in Orion}",
      journal = {arXiv e-prints},
         year = 2026,
        month = feb,
          eid = {arXiv:2602.22048},
        pages = {arXiv:2602.22048},
          doi = {10.48550/arXiv.2602.22048},
archivePrefix = {arXiv},
       eprint = {2602.22048},
 primaryClass = {astro-ph.SR},
       adsurl = {https://ui.adsabs.harvard.edu/abs/2026arXiv260222048D},
         note = {[Paper I]}
}

@ARTICLE{Duchene2013,
       author = {{Duch{\^e}ne}, Gaspard and {Kraus}, Adam},
        title = "{Stellar Multiplicity}",
      journal = {\araa},
         year = 2013,
        month = aug,
       volume = {51},
       number = {1},
        pages = {269-310},
          doi = {10.1146/annurev-astro-081710-102602},
archivePrefix = {arXiv},
       eprint = {1303.3028},
 primaryClass = {astro-ph.SR},
       adsurl = {https://ui.adsabs.harvard.edu/abs/2013ARA&A..51..269D}
}

@ARTICLE{siess2000,
       author = {{Siess}, L. and {Dufour}, E. and {Forestini}, M.},
        title = "{An internet server for pre-main sequence tracks of low- and intermediate-mass stars}",
      journal = {\aap},
         year = 2000,
        month = jun,
       volume = {358},
        pages = {593-599},
          doi = {10.48550/arXiv.astro-ph/0003477},
archivePrefix = {arXiv},
       eprint = {astro-ph/0003477},
 primaryClass = {astro-ph},
       adsurl = {https://ui.adsabs.harvard.edu/abs/2000A&A...358..593S}
}

@ARTICLE{pisa2011,
       author = {{Tognelli}, E. and {Prada Moroni}, P.~G. and {Degl'Innocenti}, S.},
        title = "{The Pisa pre-main sequence tracks and isochrones. A database covering a wide range of Z, Y, mass, and age values}",
      journal = {\aap},
         year = 2011,
        month = sep,
       volume = {533},
          eid = {A109},
        pages = {A109},
          doi = {10.1051/0004-6361/200913913},
archivePrefix = {arXiv},
       eprint = {1107.2318},
 primaryClass = {astro-ph.SR},
       adsurl = {https://ui.adsabs.harvard.edu/abs/2011A&A...533A.109T}
}

@ARTICLE{schertel2003,
       author = {{Schertl}, D. and {Balega}, Y.~Y. and {Preibisch}, Th. and {Weigelt}, G.},
        title = "{Orbital motion of the massive multiple stars in the Orion Trapezium}",
      journal = {\aap},
         year = 2003,
        month = apr,
       volume = {402},
        pages = {267-275},
          doi = {10.1051/0004-6361:20030225},
       adsurl = {https://ui.adsabs.harvard.edu/abs/2003A&A...402..267S}
}

@ARTICLE{gravity2018,
       author = {{GRAVITY Collaboration} and {Karl}, Martina and {Pfuhl}, Oliver and {Eisenhauer}, Frank and {Genzel}, Reinhard and {Grellmann}, Rebekka and {Habibi}, Maryam and {Abuter}, Roberto and {Accardo}, Matteo and {Amorim}, Ant{\'o}nio and {Anugu}, Narsireddy and {{\'A}vila}, Gerardo and {Benisty}, Myriam and {Berger}, Jean-Philippe and {Blind}, Nicolas and {Bonnet}, Henri and {Bourget}, Pierre and {Brandner}, Wolfgang and {Brast}, Roland and {Buron}, Alexander and {Caratti O Garatti}, Alessio and {Chapron}, Fr{\'e}d{\'e}ric and {Cl{\'e}net}, Yann and {Collin}, Claude and {Coud{\'e} Du Foresto}, Vincent and {de Wit}, Willem-Jan and {de Zeeuw}, Tim and {Deen}, Casey and {Delplancke-Str{\"o}bele}, Fran{\c{c}}oise and {Dembet}, Roderick and {Derie}, Fr{\'e}d{\'e}ric and {Dexter}, Jason and {Duvert}, Gilles and {Ebert}, Monica and {Eckart}, Andreas and {Esselborn}, Michael and {F{\'e}dou}, Pierre and {Finger}, Gert and {Garcia}, Paulo and {Garcia Dabo}, Cesar Enrique and {Garcia Lopez}, Rebeca and {Gao}, Feng and {Gendron}, {\'E}ric and {Gillessen}, Stefan and {Gont{\'e}}, Fr{\'e}d{\'e}ric and {Gordo}, Paulo and {Gr{\"o}zinger}, Ulrich and {Guajardo}, Patricia and {Guieu}, Sylvain and {Haguenauer}, Pierre and {Hans}, Oliver and {Haubois}, Xavier and {Haug}, Marcus and {Hau{\ss}mann}, Frank and {Henning}, Thomas and {Hippler}, Stefan and {Horrobin}, Matthew and {Huber}, Armin and {Hubert}, Zoltan and {Hubin}, Norbert and {Jakob}, Gerd and {Jochum}, Lieselotte and {Jocou}, Laurent and {Kaufer}, Andreas and {Kellner}, Stefan and {Kendrew}, Sarah and {Kern}, Lothar and {Kervella}, Pierre and {Kiekebusch}, Mario and {Klein}, Ralf and {K{\"o}hler}, Rainer and {Kolb}, Johan and {Kulas}, Martin and {Lacour}, Sylvestre and {Lapeyr{\`e}re}, Vincent and {Lazareff}, Bernard and {Le Bouquin}, Jean-Baptiste and {L{\'e}na}, Pierre and {Lenzen}, Rainer and {L{\'e}v{\^e}que}, Samuel and {Lin}, Chien-Cheng and {Lippa}, Magdalena and {Magnard}, Yves and {Mehrgan}, Leander and {M{\'e}rand}, Antoine and {Moulin}, Thibaut and {M{\"u}ller}, Eric and {M{\"u}ller}, Friedrich and {Neumann}, Udo and {Oberti}, Sylvain and {Ott}, Thomas and {Pallanca}, Laurent and {Panduro}, Johana and {Pasquini}, Luca and {Paumard}, Thibaut and {Percheron}, Isabelle and {Perraut}, Karine and {Perrin}, Guy and {Pfl{\"u}ger}, Andreas and {Duc}, Thanh Phan and {Plewa}, Philipp M. and {Popovic}, Dan and {Rabien}, Sebastian and {Ram{\'\i}rez}, Andr{\'e}s and {Ramos}, Jose and {Rau}, Christian and {Riquelme}, Miguel and {Rodr{\'\i}guez-Coira}, Gustavo and {Rohloff}, Ralf-Rainer and {Rosales}, Alejandra and {Rousset}, G{\'e}rard and {Sanchez-Bermudez}, Joel and {Scheithauer}, Silvia and {Sch{\"o}ller}, Markus and {Schuhler}, Nicolas and {Spyromilio}, Jason and {Straub}, Odele and {Straubmeier}, Christian and {Sturm}, Eckhard and {Suarez}, Marcos and {Tristram}, Konrad R.~W. and {Ventura}, Noel and {Vincent}, Fr{\'e}d{\'e}ric and {Waisberg}, Idel and {Wank}, Imke and {Widmann}, Felix and {Wieprecht}, Ekkehard and {Wiest}, Michael and {Wiezorrek}, Erich and {Wittkowski}, Markus and {Woillez}, Julien and {Wolff}, Burkhard and {Yazici}, Senol and {Ziegler}, Denis and {Zins}, G{\'e}rard},
        title = "{Multiple star systems in the Orion nebula}",
      journal = {\aap},
         year = 2018,
        month = dec,
       volume = {620},
          eid = {A116},
        pages = {A116},
          doi = {10.1051/0004-6361/201833575},
archivePrefix = {arXiv},
       eprint = {1809.10376},
 primaryClass = {astro-ph.SR},
       adsurl = {https://ui.adsabs.harvard.edu/abs/2018A&A...620A.116G}
}

@ARTICLE{kraft1967,
       author = {{Kraft}, Robert P.},
        title = "{Studies of Stellar Rotation. V. The Dependence of Rotation on Age among Solar-Type Stars}",
      journal = {\apj},
         year = 1967,
        month = nov,
       volume = {150},
        pages = {551},
          doi = {10.1086/149359},
       adsurl = {https://ui.adsabs.harvard.edu/abs/1967ApJ...150..551K}
}

@ARTICLE{alecian2013,
       author = {{Alecian}, E. and {Wade}, G.~A. and {Catala}, C. and {Grunhut}, J.~H. and {Landstreet}, J.~D. and {Bagnulo}, S. and {B{\"o}hm}, T. and {Folsom}, C.~P. and {Marsden}, S. and {Waite}, I.},
        title = "{A high-resolution spectropolarimetric survey of Herbig Ae/Be stars - I. Observations and measurements}",
      journal = {\mnras},
         year = 2013,
        month = feb,
       volume = {429},
       number = {2},
        pages = {1001-1026},
          doi = {10.1093/mnras/sts383},
archivePrefix = {arXiv},
       eprint = {1211.2907},
 primaryClass = {astro-ph.SR},
       adsurl = {https://ui.adsabs.harvard.edu/abs/2013MNRAS.429.1001A}
}

@ARTICLE{stelzer2009,
       author = {{Stelzer}, B. and {Hubrig}, S. and {Orlando}, S. and {Micela}, G. and {Mikul{\'a}{\v{s}}ek}, Z. and {Sch{\"o}ller}, M.},
        title = "{The X-ray emission from Z Canis Majoris during an FUor-like outburst and the detection of its X-ray jet}",
      journal = {\aap},
         year = 2009,
        month = may,
       volume = {499},
       number = {2},
        pages = {529-533},
          doi = {10.1051/0004-6361/200911750},
archivePrefix = {arXiv},
       eprint = {0903.4060},
 primaryClass = {astro-ph.SR},
       adsurl = {https://ui.adsabs.harvard.edu/abs/2009A&A...499..529S}
}

@ARTICLE{bagnulo2020,
       author = {{Bagnulo}, S. and {Wade}, G.~A. and {Naz{\'e}}, Y. and {Grunhut}, J.~H. and {Shultz}, M.~E. and {Asher}, D.~J. and {Crowther}, P.~A. and {Evans}, C.~J. and {David-Uraz}, A. and {Howarth}, I.~D. and {Morrell}, N. and {Munoz}, M.~S. and {Neiner}, C. and {Puls}, J. and {Szyma{\'n}ski}, M.~K. and {Vink}, J.~S.},
        title = "{A search for strong magnetic fields in massive and very massive stars in the Magellanic Clouds}",
      journal = {\aap},
         year = 2020,
        month = mar,
       volume = {635},
          eid = {A163},
        pages = {A163},
          doi = {10.1051/0004-6361/201937098},
archivePrefix = {arXiv},
       eprint = {2002.12061},
 primaryClass = {astro-ph.SR},
       adsurl = {https://ui.adsabs.harvard.edu/abs/2020A&A...635A.163B}
}

@ARTICLE{mestel1968,
       author = {{Mestel}, L.},
        title = "{Magnetic braking by a stellar wind-I}",
      journal = {\mnras},
         year = 1968,
        month = jan,
       volume = {138},
        pages = {359},
          doi = {10.1093/mnras/138.3.359},
       adsurl = {https://ui.adsabs.harvard.edu/abs/1968MNRAS.138..359M}
}

@ARTICLE{donati2009,
       author = {{Donati}, J. -F. and {Landstreet}, J.~D.},
        title = "{Magnetic Fields of Nondegenerate Stars}",
      journal = {\araa},
         year = 2009,
        month = sep,
       volume = {47},
       number = {1},
        pages = {333-370},
          doi = {10.1146/annurev-astro-082708-101833},
archivePrefix = {arXiv},
       eprint = {0904.1938},
 primaryClass = {astro-ph.SR},
       adsurl = {https://ui.adsabs.harvard.edu/abs/2009ARA&A..47..333D}
}

@ARTICLE{gregory2012,
       author = {{Gregory}, S.~G. and {Donati}, J. -F. and {Morin}, J. and {Hussain}, G.~A.~J. and {Mayne}, N.~J. and {Hillenbrand}, L.~A. and {Jardine}, M.},
        title = "{Can We Predict the Global Magnetic Topology of a Pre-main-sequence Star from Its Position in the Hertzsprung-Russell Diagram?}",
      journal = {\apj},
         year = 2012,
        month = aug,
       volume = {755},
       number = {2},
          eid = {97},
        pages = {97},
          doi = {10.1088/0004-637X/755/2/97},
archivePrefix = {arXiv},
       eprint = {1206.5238},
 primaryClass = {astro-ph.SR},
       adsurl = {https://ui.adsabs.harvard.edu/abs/2012ApJ...755...97G}
}

@ARTICLE{costero2006,
       author = {{Costero}, R. and {Echevarria}, J. and {Richer}, M.~G. and {Poveda}, A. and {Li}, W.},
        title = "{theta\^1 Orionis E}",
      journal = {\iaucirc},
         year = 2006,
        month = feb,
       volume = {8669},
        pages = {2},
       adsurl = {https://ui.adsabs.harvard.edu/abs/2006IAUC.8669....2C}
}

@ARTICLE{morales2012,
       author = {{Morales-Calder{\'o}n}, M. and {Stauffer}, J.~R. and {Stassun}, K.~G. and {Vrba}, F.~J. and {Prato}, L. and {Hillenbrand}, L.~A. and {Terebey}, S. and {Covey}, K.~R. and {Rebull}, L.~M. and {Terndrup}, D.~M. and {Gutermuth}, R. and {Song}, I. and {Plavchan}, P. and {Carpenter}, J.~M. and {Marchis}, F. and {Garc{\'\i}a}, E.~V. and {Margheim}, S. and {Luhman}, K.~L. and {Angione}, J. and {Irwin}, J.~M.},
        title = "{YSOVAR: Six Pre-main-sequence Eclipsing Binaries in the Orion Nebula Cluster}",
      journal = {\apj},
         year = 2012,
        month = jul,
       volume = {753},
       number = {2},
          eid = {149},
        pages = {149},
          doi = {10.1088/0004-637X/753/2/149},
archivePrefix = {arXiv},
       eprint = {1206.6350},
 primaryClass = {astro-ph.GA},
       adsurl = {https://ui.adsabs.harvard.edu/abs/2012ApJ...753..149M}
}

@ARTICLE{walker1969,
       author = {{Walker}, M.~F.},
        title = "{Studies of extremely young clusters. V. Stars in the vicinity of the Orion nebula.}",
      journal = {\apj},
         year = 1969,
        month = feb,
       volume = {155},
        pages = {447},
          doi = {10.1086/149881},
       adsurl = {https://ui.adsabs.harvard.edu/abs/1969ApJ...155..447W}
}

@ARTICLE{hernandez2014,
       author = {{Hern{\'a}ndez}, Jes{\'u}s and {Calvet}, Nuria and {Perez}, Alice and {Brice{\~n}o}, Cesar and {Olguin}, Lorenzo and {Contreras}, Maria E. and {Hartmann}, Lee and {Allen}, Lori and {Espaillat}, Catherine and {Hernan}, Ram{\'\i}rez},
        title = "{A Spectroscopic Census in Young Stellar Regions: The {\ensuremath{\sigma}} Orionis Cluster}",
      journal = {\apj},
         year = 2014,
        month = oct,
       volume = {794},
       number = {1},
          eid = {36},
        pages = {36},
          doi = {10.1088/0004-637X/794/1/36},
archivePrefix = {arXiv},
       eprint = {1408.0225},
 primaryClass = {astro-ph.SR},
       adsurl = {https://ui.adsabs.harvard.edu/abs/2014ApJ...794...36H}
}

@ARTICLE{hillenbrand1997,
       author = {{Hillenbrand}, Lynne A.},
        title = "{On the Stellar Population and Star-Forming History of the Orion Nebula Cluster}",
      journal = {\aj},
         year = 1997,
        month = may,
       volume = {113},
        pages = {1733-1768},
          doi = {10.1086/118389},
       adsurl = {https://ui.adsabs.harvard.edu/abs/1997AJ....113.1733H}
}

@ARTICLE{Bower2003,
       author = {{Bower}, Geoffrey C. and {Plambeck}, Richard L. and {Bolatto}, Alberto and {McCrady}, Nate and {Graham}, James R. and {de Pater}, Imke and {Liu}, Michael C. and {Baganoff}, Frederick K.},
        title = "{A Giant Outburst at Millimeter Wavelengths in the Orion Nebula}",
      journal = {\apj},
         year = 2003,
        month = dec,
       volume = {598},
       number = {2},
        pages = {1140-1150},
          doi = {10.1086/379101},
archivePrefix = {arXiv},
       eprint = {astro-ph/0308277},
 primaryClass = {astro-ph},
       adsurl = {https://ui.adsabs.harvard.edu/abs/2003ApJ...598.1140B}
}

@ARTICLE{grosscheld2019,
       author = {{Gro{\ss}schedl}, Josefa Elisabeth and {Alves}, Jo{\~a}o and {Teixeira}, Paula S. and {Bouy}, Herv{\'e} and {Forbrich}, Jan and {Lada}, Charles J. and {Meingast}, Stefan and {Hacar}, {\'A}lvaro and {Ascenso}, Joana and {Ackerl}, Christine and {Hasenberger}, Birgit and {K{\"o}hler}, Rainer and {Kubiak}, Karolina and {Larreina}, Irati and {Linhardt}, Lorenz and {Lombardi}, Marco and {M{\"o}ller}, Torsten},
        title = "{VISION - Vienna survey in Orion. III. Young stellar objects in Orion A}",
      journal = {\aap},
         year = 2019,
        month = feb,
       volume = {622},
          eid = {A149},
        pages = {A149},
          doi = {10.1051/0004-6361/201832577},
archivePrefix = {arXiv},
       eprint = {1810.00878},
 primaryClass = {astro-ph.SR},
       adsurl = {https://ui.adsabs.harvard.edu/abs/2019A&A...622A.149G}
}

@ARTICLE{petit2008,
       author = {{Petit}, V. and {Wade}, G.~A. and {Drissen}, L. and {Montmerle}, T. and {Alecian}, E.},
        title = "{Discovery of two magnetic massive stars in the Orion Nebula Cluster: a clue to the origin of neutron star magnetic fields?}",
      journal = {\mnras},
         year = 2008,
        month = jun,
       volume = {387},
       number = {1},
        pages = {L23-L27},
          doi = {10.1111/j.1745-3933.2008.00474.x},
archivePrefix = {arXiv},
       eprint = {0803.2691},
 primaryClass = {astro-ph},
       adsurl = {https://ui.adsabs.harvard.edu/abs/2008MNRAS.387L..23P}
}

@ARTICLE{simon2011,
       author = {{Sim{\'o}n-D{\'\i}az}, S. and {Garc{\'\i}a-Rojas}, J. and {Esteban}, C. and {Stasi{\'n}ska}, G. and {L{\'o}pez-S{\'a}nchez}, A.~R. and {Morisset}, C.},
        title = "{A detailed study of the H ii region M 43 and its ionizing star}",
      journal = {\aap},
         year = 2011,
        month = jun,
       volume = {530},
          eid = {A57},
        pages = {A57},
          doi = {10.1051/0004-6361/201116608},
archivePrefix = {arXiv},
       eprint = {1103.3628},
 primaryClass = {astro-ph.GA},
       adsurl = {https://ui.adsabs.harvard.edu/abs/2011A&A...530A..57S}
}

@ARTICLE{preibisch1999,
       author = {{Preibisch}, Thomas and {Balega}, Yuri and {Hofmann}, Karl-Heinz and {Weigelt}, Gerd and {Zinnecker}, Hans},
        title = "{Multiplicity of the massive stars in the Orion Nebula cluster}",
      journal = {\na},
         year = 1999,
        month = dec,
       volume = {4},
       number = {7},
        pages = {531-542},
          doi = {10.1016/S1384-1076(99)00042-1},
       adsurl = {https://ui.adsabs.harvard.edu/abs/1999NewA....4..531P}
}

@ARTICLE{schultz2019,
       author = {{Shultz}, M. and {Le Bouquin}, J. -B. and {Rivinius}, Th and {Wade}, G.~A. and {Kochukhov}, O. and {Alecian}, E. and {Petit}, V. and {Pfuhl}, O. and {Karl}, M. and {Gao}, F. and {Grellmann}, R. and {Lin}, C. -C. and {Garcia}, P. and {Lacour}, S. and {MiMeS Collaboration} and {BinaMIcS Collaboration}},
        title = "{NU Ori: a hierarchical triple system with a strongly magnetic B-type star}",
      journal = {\mnras},
         year = 2019,
        month = jan,
       volume = {482},
       number = {3},
        pages = {3950-3965},
          doi = {10.1093/mnras/sty2985},
archivePrefix = {arXiv},
       eprint = {1810.13388},
 primaryClass = {astro-ph.SR},
       adsurl = {https://ui.adsabs.harvard.edu/abs/2019MNRAS.482.3950S}
}

@ARTICLE{valegrad2021,
       author = {{Valeg{\r{a}}rd}, P. -G. and {Waters}, L.~B.~F.~M. and {Dominik}, C.},
        title = "{What happened before?. Disks around the precursors of young Herbig Ae/Be stars}",
      journal = {\aap},
         year = 2021,
        month = aug,
       volume = {652},
          eid = {A133},
        pages = {A133},
          doi = {10.1051/0004-6361/202039802},
archivePrefix = {arXiv},
       eprint = {2104.14212},
 primaryClass = {astro-ph.SR},
       adsurl = {https://ui.adsabs.harvard.edu/abs/2021A&A...652A.133V}
}

@ARTICLE{rodriguez2003,
       author = {{Rodr{\'\i}guez}, Luis F. and {G{\'o}mez}, Yolanda and {Reipurth}, Bo},
        title = "{A Cluster of Compact Radio Sources in NGC 2024 (Orion B)}",
      journal = {\apj},
         year = 2003,
        month = dec,
       volume = {598},
       number = {2},
        pages = {1100-1106},
          doi = {10.1086/378953},
archivePrefix = {arXiv},
       eprint = {astro-ph/0308480},
 primaryClass = {astro-ph},
       adsurl = {https://ui.adsabs.harvard.edu/abs/2003ApJ...598.1100R}
}

@ARTICLE{skinner2003,
       author = {{Skinner}, Stephen and {Gagn{\'e}}, Marc and {Belzer}, Emily},
        title = "{A Deep Chandra X-Ray Observation of the Embedded Young Cluster in NGC 2024}",
      journal = {\apj},
         year = 2003,
        month = nov,
       volume = {598},
       number = {1},
        pages = {375-391},
          doi = {10.1086/378085},
archivePrefix = {arXiv},
       eprint = {astro-ph/0306566},
 primaryClass = {astro-ph},
       adsurl = {https://ui.adsabs.harvard.edu/abs/2003ApJ...598..375S}
}

@ARTICLE{broos2013,
       author = {{Broos}, Patrick S. and {Getman}, Konstantin V. and {Povich}, Matthew S. and {Feigelson}, Eric D. and {Townsley}, Leisa K. and {Naylor}, Tim and {Kuhn}, Michael A. and {King}, Robert R. and {Busk}, Heather A.},
        title = "{Identifying Young Stars in Massive Star-forming Regions for the MYStIX Project}",
      journal = {\apjs},
         year = 2013,
        month = dec,
       volume = {209},
       number = {2},
          eid = {32},
        pages = {32},
          doi = {10.1088/0067-0049/209/2/32},
archivePrefix = {arXiv},
       eprint = {1309.4500},
 primaryClass = {astro-ph.SR},
       adsurl = {https://ui.adsabs.harvard.edu/abs/2013ApJS..209...32B}
}

@ARTICLE{gaia2023,
       author = {{Gaia Collaboration} and {Vallenari}, A. and {Brown}, A.~G.~A. and {Prusti}, T. and {de Bruijne}, J.~H.~J. and {Arenou}, F. and {Babusiaux}, C. and {Biermann}, M. and {Creevey}, O.~L. and {Ducourant}, C. and {Evans}, D.~W. and {Eyer}, L. and {Guerra}, R. and {Hutton}, A. and {Jordi}, C. and {Klioner}, S.~A. and {Lammers}, U.~L. and {Lindegren}, L. and {Luri}, X. and {Mignard}, F. and {Panem}, C. and {Pourbaix}, D. and {Randich}, S. and {Sartoretti}, P. and {Soubiran}, C. and {Tanga}, P. and {Walton}, N.~A. and {Bailer-Jones}, C.~A.~L. and {Bastian}, U. and {Drimmel}, R. and {Jansen}, F. and {Katz}, D. and {Lattanzi}, M.~G. and {van Leeuwen}, F. and {Bakker}, J. and {Cacciari}, C. and {Casta{\~n}eda}, J. and {De Angeli}, F. and {Fabricius}, C. and {Fouesneau}, M. and {Fr{\'e}mat}, Y. and {Galluccio}, L. and {Guerrier}, A. and {Heiter}, U. and {Masana}, E. and {Messineo}, R. and {Mowlavi}, N. and {Nicolas}, C. and {Nienartowicz}, K. and {Pailler}, F. and {Panuzzo}, P. and {Riclet}, F. and {Roux}, W. and {Seabroke}, G.~M. and {Sordo}, R. and {Th{\'e}venin}, F. and {Gracia-Abril}, G. and {Portell}, J. and {Teyssier}, D. and {Altmann}, M. and {Andrae}, R. and {Audard}, M. and {Bellas-Velidis}, I. and {Benson}, K. and {Berthier}, J. and {Blomme}, R. and {Burgess}, P.~W. and {Busonero}, D. and {Busso}, G. and {C{\'a}novas}, H. and {Carry}, B. and {Cellino}, A. and {Cheek}, N. and {Clementini}, G. and {Damerdji}, Y. and {Davidson}, M. and {de Teodoro}, P. and {Nu{\~n}ez Campos}, M. and {Delchambre}, L. and {Dell'Oro}, A. and {Esquej}, P. and {Fern{\'a}ndez-Hern{\'a}ndez}, J. and {Fraile}, E. and {Garabato}, D. and {Garc{\'\i}a-Lario}, P. and {Gosset}, E. and {Haigron}, R. and {Halbwachs}, J. -L. and {Hambly}, N.~C. and {Harrison}, D.~L. and {Hern{\'a}ndez}, J. and {Hestroffer}, D. and {Hodgkin}, S.~T. and {Holl}, B. and {Jan{\ss}en}, K. and {Jevardat de Fombelle}, G. and {Jordan}, S. and {Krone-Martins}, A. and {Lanzafame}, A.~C. and {L{\"o}ffler}, W. and {Marchal}, O. and {Marrese}, P.~M. and {Moitinho}, A. and {Muinonen}, K. and {Osborne}, P. and {Pancino}, E. and {Pauwels}, T. and {Recio-Blanco}, A. and {Reyl{\'e}}, C. and {Riello}, M. and {Rimoldini}, L. and {Roegiers}, T. and {Rybizki}, J. and {Sarro}, L.~M. and {Siopis}, C. and {Smith}, M. and {Sozzetti}, A. and {Utrilla}, E. and {van Leeuwen}, M. and {Abbas}, U. and {{\'A}brah{\'a}m}, P. and {Abreu Aramburu}, A. and {Aerts}, C. and {Aguado}, J.~J. and {Ajaj}, M. and {Aldea-Montero}, F. and {Altavilla}, G. and {{\'A}lvarez}, M.~A. and {Alves}, J. and {Anders}, F. and {Anderson}, R.~I. and {Anglada Varela}, E. and {Antoja}, T. and {Baines}, D. and {Baker}, S.~G. and {Balaguer-N{\'u}{\~n}ez}, L. and {Balbinot}, E. and {Balog}, Z. and {Barache}, C. and {Barbato}, D. and {Barros}, M. and {Barstow}, M.~A. and {Bartolom{\'e}}, S. and {Bassilana}, J. -L. and {Bauchet}, N. and {Becciani}, U. and {Bellazzini}, M. and {Berihuete}, A. and {Bernet}, M. and {Bertone}, S. and {Bianchi}, L. and {Binnenfeld}, A. and {Blanco-Cuaresma}, S. and {Blazere}, A. and {Boch}, T. and {Bombrun}, A. and {Bossini}, D. and {Bouquillon}, S. and {Bragaglia}, A. and {Bramante}, L. and {Breedt}, E. and {Bressan}, A. and {Brouillet}, N. and {Brugaletta}, E. and {Bucciarelli}, B. and {Burlacu}, A. and {Butkevich}, A.~G. and {Buzzi}, R. and {Caffau}, E. and {Cancelliere}, R. and {Cantat-Gaudin}, T. and {Carballo}, R. and {Carlucci}, T. and {Carnerero}, M.~I. and {Carrasco}, J.~M. and {Casamiquela}, L. and {Castellani}, M. and {Castro-Ginard}, A. and {Chaoul}, L. and {Charlot}, P. and {Chemin}, L. and {Chiaramida}, V. and {Chiavassa}, A. and {Chornay}, N. and {Comoretto}, G. and {Contursi}, G. and {Cooper}, W.~J. and {Cornez}, T. and {Cowell}, S. and {Crifo}, F. and {Cropper}, M. and {Crosta}, M. and {Crowley}, C. and {Dafonte}, C. and {Dapergolas}, A. and {David}, M. and {David}, P. and {de Laverny}, P. and {De Luise}, F. and {De March}, R.},
        title = "{Gaia Data Release 3. Summary of the content and survey properties}",
      journal = {\aap},
         year = 2023,
        month = jun,
       volume = {674},
          eid = {A1},
        pages = {A1},
          doi = {10.1051/0004-6361/202243940},
archivePrefix = {arXiv},
       eprint = {2208.00211},
 primaryClass = {astro-ph.GA},
       adsurl = {https://ui.adsabs.harvard.edu/abs/2023A&A...674A...1G}
}

@ARTICLE{ordonez2024,
       author = {{Ord{\'o}{\~n}ez-Toro}, Jazm{\'\i}n and {Dzib}, Sergio A. and {Loinard}, Laurent and {Ortiz-Le{\'o}n}, Gisela and {Kounkel}, Marina A. and {Masqu{\'e}}, Josep M. and {Medina}, S. -N.~X. and {Galli}, Phillip A.~B. and {Dupuy}, Trent J. and {Rodr{\'\i}guez}, Luis F. and {Quiroga-Nu{\~n}ez}, Luis H.},
        title = "{Dynamical Mass of the Ophiuchus Intermediate-mass Stellar System S1 with DYNAMO-VLBA}",
      journal = {\aj},
         year = 2024,
        month = mar,
       volume = {167},
       number = {3},
          eid = {108},
        pages = {108},
          doi = {10.3847/1538-3881/ad1bd3},
archivePrefix = {arXiv},
       eprint = {2401.02885},
 primaryClass = {astro-ph.SR},
       adsurl = {https://ui.adsabs.harvard.edu/abs/2024AJ....167..108O}
}

@ARTICLE{ordonez2025a,
       author = {{Ord{\'o}{\~n}ez-Toro}, Jazm{\'\i}n and {Dzib}, Sergio A. and {Loinard}, Laurent and {Ortiz-Le{\'o}n}, Gisela and {Kounkel}, Marina A. and {Galli}, Phillip A.~B. and {Masqu{\'e}}, Josep M. and {Dupuy}, Trent J. and {Quiroga-Nu{\~n}ez}, Luis H. and {Rodr{\'\i}guez}, Luis F.},
        title = "{VLBA detections in the Oph-S1 binary system near periastron confirmation of its orbital elements and mass}",
      journal = {\mnras},
         year = 2025,
        month = apr,
       volume = {538},
       number = {3},
        pages = {1784-1788},
          doi = {10.1093/mnras/staf396},
archivePrefix = {arXiv},
       eprint = {2503.04594},
 primaryClass = {astro-ph.SR},
       adsurl = {https://ui.adsabs.harvard.edu/abs/2025MNRAS.538.1784O}
}

@ARTICLE{forbrich2021,
       author = {{Forbrich}, Jan and {Dzib}, Sergio A. and {Reid}, Mark J. and {Menten}, Karl M.},
        title = "{A VLBA Survey of Radio Stars in the Orion Nebula Cluster. I. The Nonthermal Radio Population}",
      journal = {\apj},
         year = 2021,
        month = jan,
       volume = {906},
       number = {1},
          eid = {23},
        pages = {23},
          doi = {10.3847/1538-4357/abc68e},
archivePrefix = {arXiv},
       eprint = {2011.09329},
 primaryClass = {astro-ph.SR},
       adsurl = {https://ui.adsabs.harvard.edu/abs/2021ApJ...906...23F}
}

@ARTICLE{dzib2021,
       author = {{Dzib}, Sergio A. and {Forbrich}, Jan and {Reid}, Mark J. and {Menten}, Karl M.},
        title = "{A VLBA Survey of Radio Stars in the Orion Nebula Cluster. II. Astrometry}",
      journal = {\apj},
         year = 2021,
        month = jan,
       volume = {906},
       number = {1},
          eid = {24},
        pages = {24},
          doi = {10.3847/1538-4357/abc68f},
archivePrefix = {arXiv},
       eprint = {2011.09331},
 primaryClass = {astro-ph.SR},
       adsurl = {https://ui.adsabs.harvard.edu/abs/2021ApJ...906...24D}
}

@ARTICLE{petr1998,
       author = {{Petr}, Monika G. and {Coud{\'e} du Foresto}, Vincent and
         {Beckwith}, Steven V.~W. and {Richichi}, Andrea and
         {McCaughrean}, Mark J.},
        title = "{Binary Stars in the Orion Trapezium Cluster Core}",
      journal = {\apj},
         year = 1998,
        month = jun,
       volume = {500},
       number = {2},
        pages = {825-837},
          doi = {10.1086/305751},
       adsurl = {https://ui.adsabs.harvard.edu/abs/1998ApJ...500..825P}
}

@ARTICLE{lloyd1999,
       author = {{Lloyd}, C. and {Stickland}, D.~J.},
        title = "{The Nature of the Bright Early-Type Eclipsing Binary Theta 1 Ori A = V1016 Orionis}",
      journal = {Information Bulletin on Variable Stars},
         year = 1999,
        month = nov,
       volume = {4809},
        pages = {1},
       adsurl = {https://ui.adsabs.harvard.edu/abs/1999IBVS.4809....1L}
}

@ARTICLE{palla2001,
       author = {{Palla}, Francesco and {Stahler}, Steven W.},
        title = "{Binary Masses as a Test for Pre-Main-Sequence Models}",
      journal = {arXiv e-prints},
         year = 2001,
        month = feb,
          eid = {astro-ph/0102131},
        pages = {astro-ph/0102131},
          doi = {10.48550/arXiv.astro-ph/0102131},
archivePrefix = {arXiv},
       eprint = {astro-ph/0102131},
 primaryClass = {astro-ph},
       adsurl = {https://ui.adsabs.harvard.edu/abs/2001astro.ph..2131P}
}

@ARTICLE{marschall1988,
       author = {{Marschall}, Laurence A. and {Mathieu}, Robert D.},
        title = "{Parenago 1540: A Pre-Main-Sequence Double-Lined Spectroscopic Binary near the Orion Trapezium}",
      journal = {\aj},
         year = 1988,
        month = dec,
       volume = {96},
        pages = {1956},
          doi = {10.1086/114942},
       adsurl = {https://ui.adsabs.harvard.edu/abs/1988AJ.....96.1956M}
}

@ARTICLE{bolton1998,
       author = {{Bolton}, C.~T. and {Harmanec}, P. and {Lyons}, R.~W. and {Odell}, A.~P. and {Pyper}, Diane M.},
        title = "{HD 37017 = V 1046 ORI A double-lined spectroscopic binary with a B2e He-strong magnetic primary}",
      journal = {\aap},
         year = 1998,
        month = sep,
       volume = {337},
        pages = {183-197},
       adsurl = {https://ui.adsabs.harvard.edu/abs/1998A&A...337..183B}
}

@ARTICLE{hohle2010,
       author = {{Hohle}, M.~M. and {Neuh{\"a}user}, R. and {Schutz}, B.~F.},
        title = "{Masses and luminosities of O- and B-type stars and red supergiants}",
      journal = {Astronomische Nachrichten},
         year = 2010,
        month = apr,
       volume = {331},
       number = {4},
        pages = {349},
          doi = {10.1002/asna.200911355},
archivePrefix = {arXiv},
       eprint = {1003.2335},
 primaryClass = {astro-ph.SR},
       adsurl = {https://ui.adsabs.harvard.edu/abs/2010AN....331..349H}
}

@ARTICLE{kounkel2018,
       author = {{Kounkel}, Marina and {Covey}, Kevin and {Su{\'a}rez}, Genaro and {Rom{\'a}n-Z{\'u}{\~n}iga}, Carlos and {Hernandez}, Jesus and {Stassun}, Keivan and {Jaehnig}, Karl O. and {Feigelson}, Eric D. and {Pe{\~n}a Ram{\'\i}rez}, Karla and {Roman-Lopes}, Alexandre and {Da Rio}, Nicola and {Stringfellow}, Guy S. and {Kim}, J. Serena and {Borissova}, Jura and {Fern{\'a}ndez-Trincado}, Jos{\'e} G. and {Burgasser}, Adam and {Garc{\'\i}a-Hern{\'a}ndez}, D.~A. and {Zamora}, Olga and {Pan}, Kaike and {Nitschelm}, Christian},
        title = "{The APOGEE-2 Survey of the Orion Star-forming Complex. II. Six-dimensional Structure}",
      journal = {\aj},
         year = 2018,
        month = sep,
       volume = {156},
       number = {3},
          eid = {84},
        pages = {84},
          doi = {10.3847/1538-3881/aad1f1},
archivePrefix = {arXiv},
       eprint = {1805.04649},
 primaryClass = {astro-ph.SR},
       adsurl = {https://ui.adsabs.harvard.edu/abs/2018AJ....156...84K}
}

@ARTICLE{stelzer2006,
       author = {{Stelzer}, B. and {Micela}, G. and {Hamaguchi}, K. and
         {Schmitt}, J.~H.~M.~M.},
        title = "{On the origin of the X-ray emission from Herbig Ae/Be stars}",
      journal = {\aap},
         year = "2006",
        month = "Oct",
       volume = {457},
       number = {1},
        pages = {223-235},
          doi = {10.1051/0004-6361:20065006},
archivePrefix = {arXiv},
       eprint = {astro-ph/0605590},
 primaryClass = {astro-ph},
       adsurl = {https://ui.adsabs.harvard.edu/abs/2006A&A...457..223S}
}

@INPROCEEDINGS{montmerle2005,
       author = {{Montmerle}, T. and {Wade}, G. and {Landstreet}, J. and
         {M{\`e}nard}, F. and {Grosso}, N. and {Feigelson}, E.~D.},
        title = "{Existence, Origin and Role of Magnetic Fields in Young Early-type Stars}",
    booktitle = {Protostars and Planets V Posters},
         year = "2005",
        month = "Jan",
        pages = {8112},
       adsurl = {https://ui.adsabs.harvard.edu/abs/2005prpl.conf.8112M}
}

@BOOK{thompson2017,
       author = {{Thompson}, A. Richard and {Moran}, James M. and
         {Swenson}, George W., Jr.},
        title = "{Interferometry and Synthesis in Radio Astronomy, 3rd Edition}",
         year = "2017",
          doi = {10.1007/978-3-319-44431-4},
       adsurl = {https://ui.adsabs.harvard.edu/abs/2017isra.book.....T}
}

@ARTICLE{loinard2007,
   author = {{Loinard}, L. and {Torres}, R.~M. and {Mioduszewski}, A.~J. and 
	{Rodr{\'{\i}}guez}, L.~F. and {Gonz{\'a}lez-L{\'o}pezlira}, R.~A. and 
	{Lachaume}, R. and {V{\'a}zquez}, V. and {Gonz{\'a}lez}, E.},
    title = "{VLBA Determination of the Distance to Nearby Star-forming Regions. I. The Distance to T Tauri with 0.4\% Accuracy}",
  journal = {\apj},
archivePrefix = "arXiv",
   eprint = {0708.2081},
     year = 2007,
    month = dec,
   volume = 671,
    pages = {546-554},
      doi = {10.1086/522493},
   adsurl = {http://adsabs.harvard.edu/abs/2007ApJ...671..546L}
}

@ARTICLE{dzib2010,
   author = {{Dzib}, S. and {Loinard}, L. and {Mioduszewski}, A.~J. and {Boden}, A.~F. and 
	{Rodr{\'{\i}}guez}, L.~F. and {Torres}, R.~M.},
    title = "{VLBA Determination of the Distance to Nearby Star-forming Regions. IV. A Preliminary Distance to the Proto-Herbig AeBe Star EC 95 in the Serpens Core}",
  journal = {\apj},
archivePrefix = "arXiv",
   eprint = {1003.5900},
 primaryClass = "astro-ph.SR",
     year = 2010,
    month = aug,
   volume = 718,
    pages = {610-619},
      doi = {10.1088/0004-637X/718/2/610},
   adsurl = {http://adsabs.harvard.edu/abs/2010ApJ...718..610D}
}

@ARTICLE{ortiz2017a,
   author = {{Ortiz-Le{\'o}n}, G.~N. and {Loinard}, L. and {Kounkel}, M.~A. and 
	{Dzib}, S.~A. and {Mioduszewski}, A.~J. and {Rodr{\'{\i}}guez}, L.~F. and 
	{Torres}, R.~M. and {Gonz{\'a}lez-L{\'o}pezlira}, R.~A. and 
	{Pech}, G. and {Rivera}, J.~L. and {Hartmann}, L. and {Boden}, A.~F. and 
	{Evans}, II, N.~J. and {Brice{\~n}o}, C. and {Tobin}, J.~J. and 
	{Galli}, P.~A.~B. and {Gudehus}, D.},
    title = "{The Gould{\rsquo}s Belt Distances Survey (GOBELINS). I. Trigonometric Parallax Distances and Depth of the Ophiuchus Complex}",
  journal = {\apj},
archivePrefix = "arXiv",
   eprint = {1611.06466},
 primaryClass = "astro-ph.SR",
     year = 2017,
    month = jan,
   volume = 834,
      eid = {141},
    pages = {141},
      doi = {10.3847/1538-4357/834/2/141},
   adsurl = {http://adsabs.harvard.edu/abs/2017ApJ...834..141O}
}

@ARTICLE{kounkel2017,
   author = {{Kounkel}, M. and {Hartmann}, L. and {Loinard}, L. and {Ortiz-Le{\'o}n}, G.~N. and 
	{Mioduszewski}, A.~J. and {Rodr{\'{\i}}guez}, L.~F. and {Dzib}, S.~A. and 
	{Torres}, R.~M. and {Pech}, G. and {Galli}, P.~A.~B. and {Rivera}, J.~L. and 
	{Boden}, A.~F. and {Evans}, II, N.~J. and {Brice{\~n}o}, C. and 
	{Tobin}, J.~J.},
    title = "{The Gould{\rsquo}s Belt Distances Survey (GOBELINS) II. Distances and Structure toward the Orion Molecular Clouds}",
  journal = {\apj},
archivePrefix = "arXiv",
   eprint = {1609.04041},
 primaryClass = "astro-ph.SR",
     year = 2017,
    month = jan,
   volume = 834,
      eid = {142},
    pages = {142},
      doi = {10.3847/1538-4357/834/2/142},
   adsurl = {http://adsabs.harvard.edu/abs/2017ApJ...834..142K}
}

@ARTICLE{pecaut2013,
       author = {{Pecaut}, Mark J. and {Mamajek}, Eric E.},
        title = "{Intrinsic Colors, Temperatures, and Bolometric Corrections of Pre-main-sequence Stars}",
      journal = {\apjs},
         year = 2013,
        month = sep,
       volume = {208},
       number = {1},
          eid = {9},
        pages = {9},
          doi = {10.1088/0067-0049/208/1/9},
archivePrefix = {arXiv},
       eprint = {1307.2657},
 primaryClass = {astro-ph.SR},
       adsurl = {https://ui.adsabs.harvard.edu/abs/2013ApJS..208....9P}
}

@ARTICLE{dario2010,
       author = {{Da Rio}, N. and {Robberto}, M. and {Soderblom}, D.~R. and {Panagia}, N. and {Hillenbrand}, L.~A. and {Palla}, F. and {Stassun}, K.~G.},
        title = "{A Multi-color Optical Survey of the Orion Nebula Cluster. II. The H-R Diagram}",
      journal = {\apj},
         year = 2010,
        month = oct,
       volume = {722},
       number = {2},
        pages = {1092-1114},
          doi = {10.1088/0004-637X/722/2/1092},
archivePrefix = {arXiv},
       eprint = {1008.1265},
 primaryClass = {astro-ph.GA},
       adsurl = {https://ui.adsabs.harvard.edu/abs/2010ApJ...722.1092D}
}

@ARTICLE{pinzon2021,
       author = {{Pinz{\'o}n}, Giovanni and {Hern{\'a}ndez}, Jes{\'u}s and {Serna}, Javier and {Garc{\'\i}a}, Alexandra and {Manzo-Mart{\'\i}nez}, Ezequiel and {Roman-Lopes}, Alexandre and {Rom{\'a}n-Z{\'u}{\~n}iga}, Carlos G. and {Batista}, Maria Gracia and {Ram{\'\i}rez-V{\'e}lez}, Julio and {Osorio}, Yeisson and {Avenda{\~n}o}, Ronald},
        title = "{Understanding the Angular Momentum Evolution of T Tauri and Herbig Ae/Be Stars}",
      journal = {\aj},
         year = 2021,
        month = sep,
       volume = {162},
       number = {3},
          eid = {90},
        pages = {90},
          doi = {10.3847/1538-3881/ac04ae},
archivePrefix = {arXiv},
       eprint = {2105.12884},
 primaryClass = {astro-ph.SR},
       adsurl = {https://ui.adsabs.harvard.edu/abs/2021AJ....162...90P}
}

\appendix

\end{document}